\documentclass[ reprint,
nofootinbib,
 amsmath,amssymb,
 aps,
]{revtex4-2}
\usepackage{xcolor}
\usepackage{graphicx}
 \usepackage{dcolumn}
\usepackage{bm}
\usepackage{multirow}
\usepackage{comment}
\usepackage{enumerate}
\usepackage{soul}
\usepackage{hyperref}
\newcommand{\be}{\begin{equation}}
\newcommand{\ee}{\end{equation}}
\newcommand{\beq}{\begin{equation}}
\newcommand{\eeq}{\end{equation}}
\newcommand{\ba}{\begin{aligned}}
\newcommand{\ea}{\end{aligned}}

\newcommand{\bea}{\begin{eqnarray}}
\newcommand{\eea}{\end{eqnarray}}

\newcommand{\cO}{\mathcal{O}}

\newcommand{\cK}{\mathcal{K}}
\newcommand{\cN}{\mathcal{N}}

\newcommand{\cB}{\mathcal{B}}

\newcommand{\cI}{\mathcal{I}}

\newcommand\bi{\begin{itemize}}
\newcommand\ei{\end{itemize}}

\renewcommand{\a}{{\alpha}}

\def\Im{\mathop{\mathrm{Im}}\nolimits}

\def\Re{\mathop{\mathrm{Re}}\nolimits}

\def\unit{{1\kern-.65ex {\rm l}}}
\def\1{{1\kern-.65ex {\rm l}}}

\def\ii{{\rm i}}

\begin{document}

\begin{flushright}
{\tt\normalsize ZMP-HH/26-8}\\
\end{flushright}

\title{On Quantum Obstructions in Type IIA Orientifolds}

\author{Lukas Kaufmann}
\author{Timo Weigand}
\author{Max Wiesner}
\affiliation{%
 \vspace{6pt} II. Institut f\"ur Theoretische Physik, Universit\"at Hamburg,\\ Notkestrasse 9, 22607 Hamburg, Germany
\vspace{6pt}\\ 
 Zentrum f\"ur Mathematische Physik, Universit\"at Hamburg\\
 Bundesstra\ss e 55, 20146 Hamburg, Germany \vspace{6pt}
 }

\date{\today}

\begin{abstract}
Quantum corrections can severely modify or even remove classical infinite distance limits in four-dimensional gravity theories with minimal ${\cal N}=1$ supersymmetry. In this note we study this effect for infinite
distance directions in the classical K\"ahler moduli sector of Type IIA orientifolds at fixed four-dimensional dilaton.
 We present several independent arguments why such infinite distance directions are absent at the quantum level. 
  These involve the worldsheet theory of EFT strings and the putative asymptotically massless tower of particles.
 Key insights are provided by the uplift to M-theory on $G_2$ manifolds, which allows for a unified treatment of quantum obstructions of seemingly different origin in the dual Type IIB/F-theory frame.
  Our results apply also to the Type IIA dual of infinite distance limits for which no quantum obstruction could be detected in the Type IIB frame in previous work. In conclusion, infinite distance limits in the part of the orientifold moduli space descending from a 4d ${\cal N}=2$ vector multiplet sector are only possible at the quantum level if also some of the remaining moduli are taken to infinity.

\end{abstract}

\maketitle

\section{Introduction}\label{sec:intro}

In four-dimensional theories of quantum gravity with minimal ${\cal N}=1$ supersymmetry, quantum effects significantly modify the classical theory  not only in the interior of moduli space, but also in its asymptotic regions. 
 Recently, this general phenomenon has been analysed in detail in the complex structure sector of orientifolds of Type IIB string theory and F-theory~\cite{Kaufmann:2026fli,Kaufmann:2026mha}, combining insights in the geometry of F-theory compactifications on Calabi--Yau fourfolds, the form of explicit quantum corrections to instanton actions and the properties of EFT strings \cite{Lanza:2021udy,Martucci:2022krl,Marchesano:2022axe,Martucci:2024trp,Grieco:2025bjy, Hassfeld:2025uoy,Monnee:2025ynn}.
  Almost all classes of infinite distance limits in the complex structure moduli space of Type IIB orientifolds and F-theory (with the other moduli kept finite) turned out to be obstructed at the quantum level, in one of the following two ways~\cite{Kaufmann:2026fli,Kaufmann:2026mha}:
\begin{enumerate}
    \item There may exist an infinite distance limit in the quantum moduli space; however, this asymptotic regime is not describable using the original perturbative description, here perturbative Type II string theory or supergravity in the orientifold/F-theory setup. 
    \item The infinite distance limit is removed altogether and the quantum moduli space is effectively compact in the considered direction. 
\end{enumerate}
The quantum effects responsible for such obstructions of pure Type IIB complex structure limits can involve the string coupling $g_s$ \cite{Kaufmann:2026fli} or the K\"ahler moduli \cite{Kaufmann:2026mha}. 

However, two main open questions remained: 
The first concerns the generality of the quantum obstruction. Indeed, for a particular class of infinite distance limits in the F-theory moduli space, referred to as regular-fiber type IV limits in~\cite{Kaufmann:2026mha}, the lines of argument signalling a quantum obstruction invoked in~\cite{Kaufmann:2026mha} remain inconclusive. The second question concerns the severity of the quantum obstruction. 
 An example of the second, more drastic type of quantum obstruction listed above has been discussed in~\cite{Kaufmann:2026fli}: $g_s$ corrections completely remove a so-called type II infinite distance limit of a Calabi--Yau threefold after orientifolding. The removal of this infinite distance degeneration becomes manifest in the F-theory uplift of Type IIB string theory, which geometrizes all pure $g_s$ corrections. Unlike these, the K\"ahler corrections, which are the subject of~\cite{Kaufmann:2026mha}, cannot be captured by uplifting to a non-perturbatively complete theory since to date no such theory is known. For this reason, 
  it could not be established in~\cite{Kaufmann:2026mha} if K\"ahler corrections have a  similar effect 
 of completely removing an infinite distance limit. In principle, this is merely a technical hurdle since the distinction between $g_s$ and K\"ahler corrections is not a fundamental property of 4d $\cN=1$ theories of gravity; rather it arises due to the chosen description of the theory in terms of Type IIB orientifolds/F-theory. 

In particular, there may exist dual formulations of the theory in which $g_s$ and K\"ahler obstructions have a common origin. Relatedly, $g_s$ corrections yield an obstruction 
only for so-called O-type A infinite distance limits~\cite{Kaufmann:2026fli}, in which the O7-plane divisor factorizes in codimension one because the underlying Calabi--Yau threefold degenerates. Instead, in O-type B limits $g_s$ corrections do not yield an obstruction.\footnote{In an O-type A limit, some component of the O7-plane coincides with one of the so-called double loci over which the threefold components of a stable Calabi-Yau threefold degeneration intersect. All other limits are called O-type B. See \cite{Kaufmann:2026fli} for more details.  } Like the distinction between $g_s$ and K\"ahler corrections, also the classification into O-type A and O-type B limits is just a technical trick tailored for Type IIB orientifolds and not necessarily meaningful in other duality frames. Indeed, instead of Type IIB compactified on a Calabi--Yau threefold $V$ modded out by an orientifold action compatible with O7-planes, in this work, we consider the dual Type IIA compactification on an orientifold of the mirror Calabi--Yau $\hat{V}$. This dual perspective unifies $g_s$ and K\"ahler corrections into a single class of corrections and will allow us to address the questions which remain open from the Type IIB/F-theory perspective. 

In Sections \ref{sec_IIA} and  \ref{sec:G2}
we collect the main ingredients from Type IIA orientifolds and their uplift to M-theory on  $G_2$ manifolds needed for our analysis.
 In Section \ref{sec:EFTstrings} we argue that Type IIA orientifold quotients of infinite distance limits of type IV and III ~\cite{Grimm:2018ohb,Grimm:2018cpv,Corvilain:2018lgw,Lee:2019oct} must be quantum obstructed. The arguments presented in this section are based on the observation that the candidate EFT strings \cite{Lanza:2021udy} associated with the classical limits cannot asymptote to supergravity strings in five or six dimensions in the sense of~\cite{Katz:2020ewz,Kim:2019vuc}. To reach this conclusion, geometric insights from the lift to M-theory on $G_2$ manifolds are key. By mirror symmetry this extra information also rules out type IV limits in Type IIB/F-theory, whose fate was left open in \cite{Kaufmann:2026mha}. 
 In Section \ref{sec:NoTower} we provide an independent argument why, in fact, all pure K\"ahler moduli limits, of type II, III, or IV, must be absent in the 4d ${\cal N}=1$ quantum moduli space. The point is that there is no tower of asymptotically light states in the ${\cal N}=1$ theory, which would be required by the Distance Conjecture~\cite{Ooguri:2006in} and Emergent String Conjecture~\cite{Lee:2019oct}. 
 We compare these results  with the mirror dual situation in Section \ref{sec:Duality}.
 Since neither the distinction between $g_s$ and K\"ahler corrections, nor between O-type A and O-type B limits, has a natural counter-part on the Type IIA side, we can argue that also on the Type IIB side all pure complex structure moduli limits (at fixed K\"ahler moduli) are absent in the quantum moduli space. 
 We summarise our findings in Section \ref{sec:Conclusions}.

\section{K\"ahler limits  in Type IIA Orientifolds} \label{sec_IIA}
Type IIA string theory compactified on a Calabi--Yau threefold $\hat{V}$ gives rise to an effective 4d $\cN=2$ supergravity theory. To obtain a 4d $\cN=1$ theory, we can mod out this compactification by the well-known orientifold action
\begin{equation} \label{orientifold-IIA}
\Omega_{\rm IIA}=(-1)^{F_L}\Omega_p\tilde \sigma\,,
\end{equation}
with $F_L$ the left-moving spacetime fermion number, $\Omega_p$ worldsheet parity and $\tilde\sigma$ an anti-holomorphic involution. The latter maps
\begin{equation}\label{eq:sigmatilde}
\tilde\sigma^\ast\hat{\Omega}=\hat{\bar{\Omega}} \,, \qquad \tilde\sigma^\ast \hat{J}=-\hat{J}\,,
\end{equation}
where $\hat{\Omega}$ and $\hat{J}$ are the unique (3,0)-form and K\"ahler form of $\hat{V}$, respectively.

The classical moduli space of the resulting 4d $\cN=1$ effective theory consists of two sectors~\cite{Grimm:2004ua,Grimm:2005fa}. The first sector is spanned by the complexified K\"ahler moduli
\begin{equation} \label{Kahlersurviving}
    t^i = \int_{C_i} \left(B_2 +\ii \hat J\right)\,,
\end{equation}
where $C_i$, $i=1,\dots, h^{1,1}_-(\hat{V})$,
 form a basis of the orientifold odd homology classes $H^-_2(\hat V)$, and $B_2$ is the Type IIA Kalb-Ramond two-form. In addition, the complex structure sector of the theory is determined by the complex three-form
\begin{equation}
    \Omega_c = C_3 + \ii \Re(C \hat \Omega) \,.
\end{equation}
Here, $C_3$ is the RR three-form of Type IIA and the normalization $C$ is given by 
\begin{equation}
    C = e^{-\phi_4}\left(\int_{\hat V} \hat{\Omega} \wedge \hat{\bar{\Omega}}\right)^{-1}\,.
\end{equation}
 The 4d dilaton $\phi_4$ is related to the 10d string coupling $g_{\rm IIA}$ and the string frame volume of $\hat{V}$ via 
\begin{equation}\label{eq:4d-dilaton}
    e^{-2\phi_4}= \frac{1}{g_{\rm IIA}^2} \text{vol}(\hat{V})\,. 
\end{equation} 
The moduli spanning the complex structure sector of the classical 4d $\cN=1$ moduli space are given by the expansion coefficients 
\begin{equation}\label{def:NK0}
    \Omega_c = N^K_{(0)} \alpha_K \,,
\end{equation}
with $\alpha_K$, $K\in\{1,\dots,h^{2,1}_+(\hat{V})+1\}$, a basis of the orientifold even part of the third cohomology group, $H^3_+(\hat{V})$.

Infinite distance limits in the complex structure moduli space (with all other moduli kept fixed) of Type IIB orientifolds are mirror dual to limits in the Type IIA K\"ahler moduli space at fixed 4d dilaton and at fixed $N^K_{(0)}$. Importantly, the
 $g_s$ and K\"ahler corrections obstructing such limits in Type IIB orientifolds~\cite{Kaufmann:2026fli,Kaufmann:2026mha} map to a single kind of corrections in Type IIA. These are induced by Euclidean D2-branes wrapping 3-cycles on $\hat V$ which correspond to instantons with charge vector $\mathbf{q}$ under the three-form axions and classical action 
\begin{equation}
    S_{\mathbf{q}}=q_K N^K_{(0)}\,. 
\end{equation}
The duality to Type IIB orientifolds singles out a charge $q_0$ that can be identified with the D$(-1)$ charge. Thus, the D2-brane instantons whose charge vectors satisfy $q_K=0$ for $K\neq 0$ map to D$(-1)$ instantons on the Type IIB side. All other D2-instantons are then mirror dual to D3-brane instantons, possibly with induced D$(-1)$ charge.

\section{Uplift to M-theory on \texorpdfstring{$G_2$}{G2} Manifolds}\label{sec:G2}
Before analysing the fate of the classical K\"ahler limits at the quantum level, we recall how Type IIA orientifolds lift to M-theory compactified on a 7-manifold $X_7$ with $G_2$ holonomy~\cite{Kachru:2001je}. For this uplift to be possible, $X_7$ must admit an $S^1$-fibration,
\begin{equation} \label{S1Afibration}
    S^1_A \hookrightarrow X_7 \to \cB_6\,,
\end{equation}
whose base $\cB_6$ is identified with the compactification manifold of the Type IIA orientifold. In the presence of D6-branes and O6-planes, this fibration is non-trivial: In the vicinity of a D6-brane, the local geometry corresponds to a Taub-NUT geometry~\cite{Townsend:1995kk} with the circle $S^1_A$ shrinking to zero size at the location of the D6-brane, indicating a local weakly coupled description of the theory. Instead, in the vicinity of a single O6-plane, the local geometry of $X_7$ corresponds to an Atiyah-Hitchin manifold~\cite{Atiyah:1988jp,Seiberg:1996nz}.

The $G_2$ holonomy of $X_7$ is equivalent to the existence of a (co-)closed three-form $\Phi$ on $X_7$~\cite{10.1093/oso/9780198506010.001.0001}. This three-form, together with its Hodge dual, defines calibrated 3- and 4-cycles which are referred to as associative and coassociative, respectively. The moduli of the effective 4d $\cN=1$ theory obtained from M-theory on $X_7$ correspond to the expansion coefficients of the complexified three-form 
\begin{equation}
    \Phi+ \ii C_3 = \sum_{\a=1}^{b_3(X_7)} s^\a \omega_\a \,,\quad \omega_\a \in H^3(X_7)\,, 
\end{equation}
whose real parts are volumes of associative 3-cycles. Viewed as the uplift of a Type IIA orientifold, the $s^\a$ encode both the moduli $N^K_{(0)}$ as well as the complexified K\"ahler moduli $t^i$ in addition to 6-brane open string moduli~\cite{Acharya:2000gb,Acharya:2001gy}. Geometrically, we can distinguish associative three-cycles depending on their position relative to the $S^1_A$-fiber in $X_7$. In particular, there are associative three-cycles $\Sigma_K$ that lie purely in the base such that 
\begin{equation}\label{def:sK}
    s^K \equiv \int_{\Sigma_K} (\Phi+\ii C_3)\,
\end{equation}
can classically be identified with $N^K_{(0)}$. Euclidean M2-branes wrapping $\Sigma_K$ then map to D2-brane instantons in the Type IIA orientifold. Instead, a second class of associative three-cycles $\Gamma_i$ is given by the restriction of the $S^1_A$-fibration to complex curves in $\cB_6$, i.e., 
\begin{equation}
    S^1_A \hookrightarrow \Gamma_i \to C_i\,. 
\end{equation}
The moduli 
\begin{equation}\label{def:si}
    s^i \equiv \int_{\Gamma_i} (\Phi+\ii C_3)\,
\end{equation}
classically correspond to the moduli $t^i$ of the Type IIA dual. M2-branes wrapping $\Gamma_i$ map to worldsheet instantons in the dual Type IIA description.

A particularly simple geometry is obtained if in the orientifold background $\hat{V}/\Omega_{\rm IIA}$, the O$6$-tadpole is cancelled locally. In this case, the $G_2$ manifold $X_7$ arising in the M-theory uplift is given by~\cite{10.1093/oso/9780198506010.001.0001,Kachru:2001je}
\begin{equation}\label{eq:X7}
    X_7=\left(\hat{V}\times S^1_A\right)/(\tilde\sigma,R)\,,
\end{equation}
where $R$ acts as a reflection on the $S^1_A$ coordinate $y$, see e.g.~\cite{Braun:2019wnj} for concrete examples. The three-form $\Phi$ can be expressed in terms of the original Calabi--Yau calibrations introduced around~\eqref{eq:sigmatilde} as
\begin{equation}
    \Phi=\Re(\hat{\Omega})+\hat{J}\wedge{\rm d}y\,,\quad \star\Phi=\Im(\hat{\Omega})+\frac{1}{2}\hat{J}\wedge \hat{J}\,.
\end{equation}
Holomorphic cycles in $\hat{V}$ then map to calibrated cycles in $X_7$. In particular, an orientifold even holomorphic divisor $\hat{D}$ of $\hat{V}$ maps to the coassociative 4-cycle
\begin{equation}\label{eq:DG2}
    D=\left(\hat{D}\times\{0\}\right)/(\tilde\sigma,R)\,
\end{equation}
in $X_7$. These coassociative 4-cycles play a crucial role in the following.

\section{Candidate EFT strings in Type IIA orientifolds}\label{sec:EFTstrings}

 As explained in Section \ref{sec_IIA}, we consider asymptotic regions in the ${\cal N}=1$ K\"ahler moduli space of Type IIA orientifolds where
\begin{equation}\label{limits} 
    t^i\to \ii \infty \,,\quad i\in \cI\subset\{1,\dots, h^{1,1}_-(\hat{V}/\Omega_{\rm IIA})\}\,,
\end{equation} 
at finite complex structure moduli $N^K_{(0)}$ and finite 4d dilaton. 
  These limits descend from corresponding limits in the Type IIA vector multiplet moduli space prior to orientifolding. 
 There are three types of such parent 4d ${\cal N}=2$ limits~\cite{Corvilain:2018lgw,Lee:2019oct}, which are mirror dual to complex structure limits of type II, III, and IV \cite{Grimm:2018ohb,Grimm:2018cpv}.
  These are, respectively, interpreted as emergent string limits (type II), or as decompactification limits to six (type III) or five (type IV) spacetime dimensions \cite{Lee:2019oct,Hassfeld:2025uoy,Monnee:2025ynn}.

In four-dimensional theories with at least $\cN=1$ supersymmetry, infinite distance limits are closely related to certain string solutions, dubbed EFT strings in~\cite{Lanza:2021udy} and further studied in \cite{Martucci:2022krl,Marchesano:2022axe,Cota:2022yjw,Martucci:2024trp,Grieco:2025bjy, Hassfeld:2025uoy,Monnee:2025ynn}. Their backreaction induces a profile for the moduli in spacetime that realizes the infinite distance limit at the core of the solution. In general, it is expected~\cite{Lanza:2021udy} that for a large class of infinite distance limits in the moduli space, so-called EFT string limits, such an EFT string solution exists.\footnote{In an EFT string limit, all moduli which scale to infinity do so at the same rate. The question whether or not quantum corrections modify a classical limit is not expected to depend on whether a limit is of this type or whether the scaling behaviour is more general. It is understood from now on that our analysis of limits via EFT strings strictly speaking applies to such EFT string limits. The analysis in Section \ref{sec:NoTower} holds also for putative non-EFT string limits.}

This gives a criterion for a classical EFT string limit in the K\"ahler moduli space of Type IIA orientifolds to survive at the quantum level: 
 There must exist a solitonic string in the 4d ${\cal N}=1$ theory with the expected properties of an EFT string associated with this limit.

More precisely, the candidate EFT strings associated with the 4d ${\cal N}=1$ limits are charged under the two-forms dual to the axions $\text{Re}\,t^i$ for $i\in \cI$.    
 They are therefore macroscopically realized as NS5-branes wrapping a nef and holomorphic divisor $\hat{D}$ \cite{Lanza:2021udy}.
  The divisor $\hat D$ is determined as follows: The backreaction of the string near its core is such that the volume moduli of all curves $C$ with $C \cdot \hat D > 0$ scale to infinity, as in a limit of the form \eqref{limits}.

 Compatibility with the orientifold projection requires that the divisor class of $\hat D$ is orientifold even, ${[\hat D] \in H^+_4(\hat V)}$~\cite{Lanza:2021udy}.
 To appreciate the latter point, we expand the homology class of $\hat D$ into its orientifold even and odd components, $[\hat D] = [\hat D_+] + [\hat D_-]$, with ${[\hat D_\pm] \in H_4^\pm(\hat V)}$.
  If $[\hat D_-] \neq 0$, there would exist some holomorphic curve class $[C_+] \in H_2^+(\hat V)$, invariant under the orientifold, such that $0 \neq [\hat D_-] \cdot [C_+] = [\hat D] \cdot [C_+]$; in the second step we used that due to the anti-symmetry of the volume form on $\hat V$, $[\hat D_\pm]$ has a non-trivial intersection pairing only with curve classes $[C_\mp] \in H_2^\mp(\hat V)$ of opposite parity under the involution.
 However, this would mean that in the limit some of the K\"ahler moduli of $\hat V$ which are projected out by the orientifold would scale to infinity, i.e. the limit would be incompatible with the orientifold projection. Indeed, recall from \eqref{Kahlersurviving} that the surviving K\"ahler moduli are associated with the orientifold odd curve classes.  
 
 Suppose now that the class $[\hat D] = [\hat D_+]$ has a smooth, irreducible, reduced representative $\hat D$. 
 A single NS5-brane can then be wrapped on $\hat D$ and gives rise to a candidate EFT string.\footnote{ Note that this scenario can arise by starting from a smooth divisor on $\hat V$ which is invariant as a whole under $\tilde\sigma$, or by taking the union $\hat D = D' + \tilde\sigma(D')$ of a divisor $D'$ and its orientifold image and deforming it into a smooth irreducible divisor.}
 As the magnetic dual of the Kalb-Ramon form, $B_6$, is even under $\Omega_{\rm IIA}$, the NS$5$-brane on the orientifold even divisor $\hat D$ is mapped to itself by $\Omega_{\rm IIA}$. Hence, this EFT string survives the orientifold projection as a BPS object.
 In the orientifolded theory, this string is the candidate EFT string for the limit in the classical $\cN=1$ moduli space in which the volume moduli of all curve classes $[C_-] \in H^-_2(\hat V)$ with positive intersection with $[\hat D]$ scale to infinity.
 
As we will explain in Section \ref{subsec:smooth}, for orientifolds of type IV and type III limits 
a smooth, irreducible and reduced EFT string divisor $\hat D$ is guaranteed to exist. For type II limits, on the other hand, a subtlety~\cite{Lanza:2021udy} related to the quantisation of the minimal integral basis of $H^2_+(\hat V)$ might imply that $\hat D$ is not reduced.
 
 For now let us focus on limits of type III and IV.
 As recalled above, at the  level of the classical effective action, these must describe decompactification limits to six and five dimensions, respectively. For details of this reasoning, in the mirror dual Type IIB framework, we refer to \cite{Monnee:2025ynn} and references therein.
 Correspondingly, the underlying EFT string must admit an interpretation as a gravitationally coupled string in six or five dimensions. 
 However,  if we take into account the effects of the orientifold projection, we will see that such an interpretation of the candidate EFT string is incompatible with the  properties of its worldsheet theory. 
  This signals a quantum obstruction for these limits. This is similar in spirit, and in fact complementary, to the strategy of \cite{Kaufmann:2026mha}. 
  There, the candidate EFT string perspective was used to argue for quantum obstructions to the type II and III cases on the dual Type IIB side. 
  However, these arguments remained inconclusive for limits of type IV. 
  As we will now explain, for the mirror dual Type IIA 
  limits, the lift to M-theory on $G_2$ manifolds provides additional information which points to quantum obstructions also for orientifolds of type IV limits.

\subsection{EFT string candidates and \texorpdfstring{$G_2$}{G2} uplifts}\label{G2uplifts}
Consider first a classical type IV limit.
In the parent 4d ${\cal N}=2$ theory, such a limit is a decompactification limit for the M-theory circle $S^1_A$ \cite{Corvilain:2018lgw, Lee:2019oct}. 
 This follows from the fact that the limit is taken at constant 4d dilaton (\ref{eq:4d-dilaton}), while in a Type IV limit the K\"ahler moduli scale to infinity in such a way that the string scale volume ${\rm vol}(\hat V)$ diverges. This requires a rescaling of $g_{\rm IIA}$, which is identified with the radius of $S^1_A$.

 In the orientifold theory, the fate of the candidate EFT string, and hence of the limit, is most conveniently understood by studying the lift to M-theory compactified on a $G_2$ manifold
$X_7=(\hat{V}\times S^1_A)/(\tilde{\sigma},R)$, which was introduced around~\eqref{eq:X7}. 
 The candidate EFT string obtained from an NS5-brane on the divisor $\hat D \subset \hat V$ lifts to an M5-brane on a coassociative 4-cycle $D\subset X_7$, which is constructed from $\hat D$ as in \eqref{eq:DG2}. Importantly, this 4-cycle is localised at $y=0$ or $y=\pi$, with $y$ the coordinate along $S^1_A$. 
 In other words, the orientifold action projects out the real scalar parametrising the motion along the M-theory circle. This only occurs if $\hat D$ is by itself a smooth, irreducible, and reduced divisor on $\hat V$ such that the orientifold action maps the spectrum on a single NS5-brane along $\hat D$ to itself. As anticipated already and shown in Section \ref{subsec:smooth}, this is always the case for type IV limits.\footnote{If, by contrast, $\hat D$ were not irreducible and reduced because it necessarily had to split into two components $D'$ and $\tilde \sigma(D')$, the NS5-branes on each of them would be exchanged by the orientifold projection. In this case, there would remain one combination of scalar fields in the spectrum which parametrises the motion of the brane-image brane pair along the M-theory circle. } 

In such situations, we therefore see that $D$ cannot move freely along the $S^1_A$-fiber of $X_7$ so that the candidate EFT string obtained from the M5-brane on $D$ cannot propagate along the $S^1_A$ direction. This indicates that the candidate EFT string cannot correspond to a supergravity string, in the sense of~\cite{Katz:2020ewz}, of an asymptotically five-dimensional theory that is obtained by decompactifying $S^1_A$:  Such a string would have to be freely moving along all (asymptotically) macroscopic spacetime dimensions, which in particular include the growing direction along $S^1_A$. Thus, the candidate EFT string is manifestly incompatible with the classical type IV limit. What we can conclude at this stage is that the limit does either not exist at all in the quantum moduli space, or at least that the asymptotic physics differs considerably from the classical expectations. Either way the limit is quantum obstructed in the sense explained in the Introduction. We will, in fact, argue for the first option in Section \ref{sec:NoTower}.

One might wonder if the conclusions are invalidated if the generically irreducible EFT string divisor $\hat D$ can split into two components in higher codimension along its moduli space, $\hat D \to D' \cup \tilde\sigma(D')$; both components could then move together as a pair such that their positions along $S^1_A$ are mapped to each other by the orientifold reflection $R$. However, the quantum worldsheet theory on the string always probes its full scalar field moduli space~\cite{Coleman:1973ci} and therefore an inconsistency along generic points in moduli space cannot be avoided by restricting to a higher codimension sublocus by hand.

Note that the argument is not restricted to the M-theory uplift of orientifolds with local tadpole cancellation, as given in~\eqref{eq:X7}, but also holds in the context of more general $G_2$ manifolds with the structure of a genuine $S^1_A$ fibration \eqref{S1Afibration}.

Furthermore, the same reasoning applies to candidate EFT strings of orientifolds of type III limits, which must describe decompactifications to six dimensions: The EFT string is fixed along the $S^1_A$ fiber of the M-theory uplift and cannot correspond to a gravitational string moving freely in the asymptotically macroscopic six spacetime dimensions. 
By contrast, the same arguments do not carry over straightforwardly to orientifolds of type II limits as for these, as explained in Section \ref{subsec:smooth}, the existence of a smooth, irreducible, reduced divisor $\hat D$ is less clear.
However, for such limits the mirror dual Type IIB picture \cite{Kaufmann:2026mha} already provides an explicit EFT string based argument why these limits are quantum obstructed.

\subsection{Zero mode analysis} 

We can confirm that the deformation mode along $S^1_A$ is indeed projected out by analysing the worldsheet spectrum of the EFT string, more closely paralleling the mirror dual procedure in~\cite{Kaufmann:2026mha}. 

 On a $G_2$ manifold $X_7$, the zero modes on the M$5$-brane string wrapped on a coassociative 4-cycle $D$ are counted as follows. First, there are two scalars corresponding to the normal direction of the string in the extended spacetime. Next, the number of (real) internal deformations of $D$ inside $X_7$ is given by $b^2_+(D)$~\cite{Mclean}, where the subscript refers to self-duality on $D$. Finally, the reduction of the anti-self-dual worldvolume two-form of the M$5$-brane contributes $b^2_+(D)$ right-moving modes. 
 
 Consider first the seven-manifold $\tilde{X}_7=\hat{V}\times S^1_A$, which is simply the M-theory uplift of Type IIA on $\hat{V}$ prior to orientifolding. On $\tilde{X}_7$, we can consider the co-associative 4-cycle $\tilde{D} = \hat{D}\times \{\rm pt\}$, where $\rm pt$ is a point along $S^1_A$ and $\hat{D}$ is the smooth, irreducible and reduced divisor in $\hat{V}$ discussed above. An M5-brane wrapping $\tilde{D}$ is the M-theory uplift of the EFT string of the 4d $\cN=2$ parent theory and corresponds to an MSW string~\cite{Maldacena:1997de}. For the 4-cycle $\tilde{D}$, we have 
 \begin{equation}
     b^2_+(\tilde{D}) = h^{2,0}(\hat{D}) + h^{0,2}(\hat{D}) + h^{1,1}_+(\hat{D})\,,
 \end{equation}
 where $h^{1,1}_+(\hat{D})$ denotes the self-dual part of $h^{1,1}(\hat{D})$. Since 
 \begin{equation}
     {\rm sgn}\;H^{1,1}(\hat{D}) = (1,\rho)\,,\quad \rho\geq 0\,,
 \end{equation}
 the single element in $H^{1,1}_+(\hat{D})$ gives rise to 
 one real deformation mode of $\tilde{D}$ that is paired with a single right-moving mode from the anti-self-dual two-form reduced over the element in $H^{1,1}_+(\hat{D})$. Whereas the deformation modes of $\tilde{D}$ counted by $h^{2,0}(\hat{D})$ and $h^{0,2}(\hat{D})$ together describe the (complex) deformations of $\hat{D}$ inside $\hat{V}$, the single real mode counted by $h^{1,1}_+(\hat{D})=1$ describes the motion of $\tilde{D}$ in $S^1_A$. Altogether, for the $\cN=2$ EFT string obtained from an M5-brane on $\tilde{D}$ we have
 \begin{equation}
    \tilde c_R= 6 \left(h^{2,0}(\hat{D}) +1\right)\,,
\end{equation}
where the prefactor arises due to the fermions that complete the right-moving scalar modes into full 2d $\cN=(0,4)$ multiplets. 

We can now consider the effect of the orientifold projection, which at the level of the seven-fold $\tilde{X}_7$ amounts to modding out by $(\tilde{\sigma},R)$ as in~\eqref{eq:X7}. This maps the 4-cycle $\tilde{D}$ to $D$ defined in~\eqref{eq:DG2}. The construction of $D$ makes it evident that the self-dual two-forms on $D$ are precisely the $\tilde\sigma$-even self-dual two-forms on $\hat{D}$. Correspondingly, we find that there are $b^2_+(D)=h^{2,0}(\hat{D})$ real zero modes on the M$5$-brane string corresponding to internal deformation modes of $D$ inside $\hat{V}/\tilde{\sigma}$. Importantly, the unique self-dual element of $H^{1,1}(\hat{D})$ is anti-invariant under $\tilde{\sigma}$ because it corresponds to the K\"ahler form on $\hat D$, which is projected out by the symplectic involution, see~\eqref{eq:sigmatilde}. The corresponding deformation mode, which would have described the position of $D$ along $S^1_A$, is therefore projected out from the worldsheet of the M5-brane on~$D$. This reflects the fact that the 4-cycle $D$ is not movable along $S^1_A$. Taking into account the right-moving superpartners of these bosonic zero modes, we therefore obtain 
\begin{equation}\label{eq:cR-orientifold}
    {c}_R = 3 \left(h^{2,0}(\hat{D})+1\right)=\frac12 \tilde c_R\,. 
\end{equation}
It is instructive to compare the situation to the mirror dual analysis of~\cite{Kaufmann:2026mha}, 
 which argued that certain infinite distance limits in Type IIB orientifolds/F-theory are 
quantum obstructed by showing that the right-moving central charge of their associated candidate EFT string is inconsistent with the prediction of the classical effective action. In particular, $c_R$ does not correspond to the value required for a critical string (for type II limits) or a 6d supergravity string as defined in~\cite{Kim:2019vuc} (for type III limits), contrary to the asymptotic physical interpretation of the limits \cite{Monnee:2025ynn}. For type IV limits, in which the candidate EFT string has to correspond to a 5d supergravity string (in the sense of~\cite{Katz:2020ewz}), a quantum obstruction could not be argued for based on $c_R$ alone. Here, we face the same problem since the right-moving central charge, $c_R$, computed in~\eqref{eq:cR-orientifold} could in principle correspond to the central charge of a 5d supergravity string. 

In the F-theory analysis of~\cite{Kaufmann:2026mha}, there is no clear way to circumvent this problem since this would require access to the $\alpha'$-completed uplift of F-theory itself. Compared to F-theory on elliptic Calabi--Yau fourfolds, M-theory compactifications on $G_2$ manifolds have a crucial advantage since in this formulation of the 4d $\cN=1$ theory, the geometry of $X_7$ encodes even more information than the fourfold in F-theory. 
Indeed, the rigidity of the uplift of the candidate EFT string to M-theory is exactly the type of inconsistency which was not manifest for type IV limits in the mirror dual Type IIB/F-theory setup.

\subsection{Condition for smoothness} \label{subsec:smooth}

Finally, as promised in Section \ref{G2uplifts}, we now explain why for orientifolds of type IV or III limits a smooth, irreducible and reduced EFT string divisor $\hat D$ always exists and which subtleties could arise for type II limits.

 Recall first that on a Calabi--Yau threefold $\hat{V}$
  general theorems guarantee 
  the existence of a smooth and irreducible representative of a nef and holomorphic divisor class $[\hat{D}]$  (or a multiple thereof) with extra properties.
  A nef and holomorphic divisor class $[\hat{D}]$ is automatically semi-ample~\cite{Oguiso},
 meaning that the linear system $|n\hat{D}|$ is basepoint free for $n\in\mathbb{N}$ large enough. By Bertini's smoothness theorem, see e.g.~\cite{Iitaka}, the generic member of $|n\hat{D}|$ is smooth, but in general not irreducible. To see which extra conditions are needed for irreducibility, we recall that the basepoint free linear system $|n\hat{D}|$ defines a holomorphic map
\begin{equation}
    \varphi_{|n\hat{D}|}:\hat{V}\to\mathbb{P}^N\,,\quad N=h^0(\hat{V},n\hat{D})-1\,,
\end{equation}
by taking a basis of sections of $n\hat{D}$ as projective coordinates. Bertini's connectedness theorem~\cite{lazarsfeld} states that the generic member of $|n\hat{D}|$ is irreducible if and only if the image of $\varphi_{|n\hat{D}|}$ has dimension at least two.

For the study of large volume infinite distance limits in the K\"ahler sector of Type IIA on the orientifold $\hat{V}/\Omega_{\rm IIA}$, 
 the divisor class $[\hat D]$ associated with the corresponding EFT string is, by construction, nef and holomorphic~\cite{Lanza:2021udy}.
 The property to check is therefore irreducibility of the generic member of $|n\hat{D}|$. For limits of type IV, the associated divisor $\hat{D}$ is big, i.e. $\hat{D}^3>0$ \cite{Corvilain:2018lgw,Lee:2019oct,Lanza:2021udy}. This implies that $\dim(\Im(\varphi_{|n\hat{D}|}))=\dim\hat{V}=3$ so that we can indeed assume the divisor underlying a type IV limit to be both smooth and irreducible. In particular, $|n\hat{D}|$ must have at least $N+1=4$ independent sections. Similarly, for type III limits $\hat{V}$ is elliptically fibered over some base $B_2$  and the divisor $\hat{D}=\pi^\ast\hat{C}$ is necessarily vertical over a holomorphic and big curve $\hat{C}$, $\hat{C}^2>0$, on the base $B_2$ \cite{Lee:2019oct, Lanza:2021udy}. In this case, the map $\varphi_{|n\hat{D}|}$ factors through the base $B_2$ and thus has image of dimension two, i.e. $N+1\geq3$.

Note that these key properties of $\hat D$ do not change for EFT strings of orientifolds of type IV or III limits: Even though in these cases, $\hat D \cdot C_+ =0$ for all orientifold even curve classes\footnote{In particular, for $h^{1,1}_+(\hat{V})\neq0$, $\hat{D}$ necessarily lies on the boundary of the full K\"ahler cone of $\hat{V}$ and can therefore not be ample.}, it still holds that $\hat D^3 > 0$ for type IV limits and $\hat D^2>0$ for $\hat{D}=\pi^\ast\hat{C}$ with $\hat{C}^2>0$ on $B_2$ in type III limits.

The situation changes for limits of type II: For these $\hat{V}$ has to allow for a surface fibration with fiber class $F$~\cite{Lee:2019oct}. While the class $F$ itself is smooth and irreducible, multiples $nF$ with $n>1$ can never be irreducible. For type II limits in 4d ${\cal N}=2$ theories, this is immaterial because the EFT string divisor can be taken to be $F$. However, the orientifold introduces a subtlety: The quantisation of $H^2_+(\hat{V},\mathbb{Z})$ might require that the correct candidate EFT string for a type II limit is associated with a multiple $nF$ with $n =2$, as explained in \cite{Lanza:2021udy}.\footnote{In fact, a similar quantisation issue could force us to replace $\hat D$ by $2\hat D$ also for orientifolds of type IV or III limits, but this poses no problem because the theorems described above still guarantee the existence of an irreducible and reduced divisor in this class.} In this case, no irreducible, reduced divisor $\hat D$ exists, and correspondingly the scalar mode parametrising the position of the M5-brane along the M-theory circle is not projected out. This seems to be prevent us from running a similar argument based on the localisation of the EFT string along $S^1_A$ as for type III and IV EFT strings.
We reiterate, however, that despite this technicality, for type II limits the EFT string perspective in the mirror dual Type IIB framework already points to a quantum obstruction of the limit~\cite{Kaufmann:2026mha}.

\section{Absence of towers of light states} \label{sec:NoTower}

A quantum obstruction for large (string frame) volume limits at constant 4d dilaton in Type IIA orientifolds can also be inferred without invoking candidate EFT strings. Namely, we can use the Distance and Emergent String Conjectures~\cite{Ooguri:2006in, Lee:2019oct} to argue that such a classical infinite distance limit is obstructed because there is no tower of asymptotically light states.  
This approach thus reverses the logic compared to typical tests of these conjectures as reviewed, e.g., in~\cite{vanBeest:2021lhn}.

Before the orientifold projection, 
the towers of states that become light in the K\"ahler limits include
the tower of D$0$-branes. Depending on the specific type of limit, the D0-branes are interpreted as the KK tower of a decompactification along a growing $S^1$ to 5d (type IV), to 6d (type III) or as the KK part of the tower of excitations of an emergent heterotic or Type II string compactified on a manifold containing an $S^1$-factor (type II) \cite{Lee:2019oct}. 
  In the 4d ${\cal N}=1$ theory, however,
  a D$0$-brane is not invariant under the orientifold action. Rather, it is mapped to an anti D$0$-brane. On the orientifolded background we can hence consider at best a tower of D$0$-$\overline{{\rm D}0}$ states. These are non-BPS and, in fact, unstable due to a tachyonic mode between the D$0$- and the $\overline{{\rm D}0}$-brane. This holds irrespective of the details of the orientifold involution or the specific scalings of the K\"ahler moduli of $\hat V$ which distinguish between limits of type II, III or IV.

Even though the D$0$-$\overline{{\rm D}0}$ states are non-BPS and unstable, one might wonder whether they become massless in the ${\rm vol}(\hat{V})\sim g_{\rm IIA}^2\to \infty$ limit. In the following, we give three reasons from different perspectives why this cannot be the case: 
\begin{enumerate}[I.]
    \item First, the mass of a non-BPS state is not protected from perturbative or non-perturbative quantum corrections. In particular, the self-energy of the D$0$-brane is expected to induce a mass correction of the form
\begin{equation} \label{eq:D0mass}
    \frac{m_{{\rm D}0\text{-}\overline{{\rm D}0}}}{M_s}=\frac{2 \pi}{g_{\rm IIA}}\left(1+ag_{\rm IIA}+\cO(g_{\rm IIA}^2)\right)\,.
\end{equation}
 Since the 4d dilaton~\eqref{eq:4d-dilaton} is kept constant in the limit, this means that also the D$0$-$\overline{{\rm D}0}$ mass in Planck units is expected to receive ${\cal O}(1)$ corrections in the limit. 
  While the computation of the precise prefactor $a$ of the correction \eqref{eq:D0mass} is beyond the scope of this work, in absence of BPS protection a perturbative renormalisation of the mass is generic, and in the limit $g_{\rm IIA} \to \infty$ leads to a sizeable correction. 

\item There is an indirect argument why the coefficient $a$ does not vanish identically. If the mass of the  D$0$-$\overline{{\rm D}0}$-states was uncorrected, it would still be among the lightest towers in the $g_{\rm IIA}\to \infty$ limit. The Emergent String Conjecture \cite{Lee:2019oct} then requires the  D$0$-$\overline{{\rm D}0}$-tower to either be a KK tower of a dual theory or part of the excitation spectrum of an emergent, weakly coupled string. However, in the large radius limit, KK towers have to be sufficiently long-lived and cannot correspond to states with a tachyonic instability. Similarly, the physical states of a critical string become asymptotically stable in the weak-coupling limit. 
  
As a consequence, the unstable  D$0$-$\overline{{\rm D}0}$-system cannot furnish the tower of states predicted by the Emergent String Conjecture. In particular, this means that there is no  decompactification limit to a 5d theory of gravity, unlike what one would have classically expected for a type IV limit. This is consistent with the candidate EFT string for this limit not corresponding to a 5d supergravity string, as concluded in the previous section.
  
\item Finally, the absence of a tower of light D$0$-$\overline{{\rm D}0}$ states can be inferred from the perspective of M-theory compactified on a $G_2$ manifold $X_7$. Given the definition of the M-theory moduli $s^i$ and $s^K$ in~\eqref{def:si} and~\eqref{def:sK}, respectively, the limits under consideration amount to taking $s^i\to \infty$ for $i\in \cI$ at constant $s^K$. In presence of D$6$-branes and O$6$-planes, the $S^1_A$-fibration of $X_7$ is non-trivial so that $X_7$ has no isometry and the momentum of a KK-mode along $S^1_A$ is not protected by such a symmetry. Instead, a KK-momentum along $S^1_A$ can flow out to the base and is hence not confined to $S^1_A$.\footnote{Of course there are the overall KK-modes of the full 7-manifold that also include $S^1_A$ but these propagate along the entire 7-manifold and are not restricted to the $S^1_A$ fiber.}

\end{enumerate} 
These three perspectives consistently show that there is no tower of light states in the classical M-theory limit of Type IIA orientifolds that would play the role of a KK tower in a 5d decompactification limit.

In the absence of the KK-tower dual to D0-branes, one might consider the tower of Type IIA supergravity KK-modes along $\hat{V}/\Omega_{\rm IIA}$ as the tower of light states predicted by the Distance Conjecture~\cite{Ooguri:2006in}. Indeed, these modes become light in Type IIA string units as 
\begin{equation}
    \frac{M_{{\rm KK}, \hat{V}}}{M_s} \sim g_{\rm IIA}^{-1/3} \to 0\,.  
\end{equation}
However, since in this regime the theory is best described by M-theory on $X_7$, the relevant UV scale is given by the fundamental M-theory scale $M_{11}$. The aforementioned KK-modes now correspond to the supergravity KK-modes of the base $\cB_6$ of $X_7$ in~\eqref{S1Afibration}. Since the M-theory volume of the base remains constant in the limit, the tower of KK-modes associated with $\cB_6$ does not become massless in M-theory units. Hence, the M-theory analysis suggests that there is no tower of states arising in the limit. This, in turn, indicates that the classical infinite distance limit itself is absent in the full theory and the corresponding direction in the quantum moduli space is compact.\footnote{This is reminiscent of compactifications of the heterotic $E_8\times E_8$ string to four dimensions, where the decompactification limit to Ho\v{r}ava--Witten theory is obstructed~\cite{Witten:1996mz,Cvetic:2024wsj} and no infinite distance limit remains at the quantum level~\cite{Cvetic:2025nfx}.}

The absence of the classical infinite distance limit ${s^i\to \infty}$ at constant $s^K$ has implications for the structure of the moduli space of M-theory compactified on an $S^1_A$-fibered $G_2$ manifold. Geometrically, this regime corresponds to the non-adiabatic limit of the fibration in which the $S^1_A$-fiber becomes large while the base volume remains constant. By our arguments above, this regime does not correspond to an infinite distance limit due to the absence of a tower of light states. For this reason, this regime of the theory cannot be described within the supergravity approximation. This, in turn, implies that there are quantum corrections that become unsuppressed in this regime. Natural candidates for these quantum corrections are one-loop corrections to the action of M2-brane instantons on 3-cycles or threshold corrections to the gauge kinetic function on D6-branes. For M2-brane instantons on 3-cycles $\Sigma_K$ the classical instanton action is given by $s^K$. As also discussed for the dual D3-brane instantons in Type IIB/F-theory in~\cite{Kaufmann:2026mha}, a quantum obstruction to the $s^i\to \ii \infty$ limit is manifest if there is a 1-loop correction of the form
\begin{equation}\label{eq:Sq} 
    S_{\mathbf{q}}=q_K s^K+b_{\mathbf{q}}(p_is^i)^{\alpha_{\mathbf{q}}}\,,\quad  \alpha_{\mathbf{q}}>0\,,\;b_{\mathbf{q}}\in \mathbb R\,,
\end{equation}
or, equivalently, a threshold correction to the gauge kinetic function on D6-branes. The latter has been computed for toroidal orbifolds in~\cite{Lust:2003ky,Akerblom:2007np,Blumenhagen:2007ip}.

To remain within the supergravity approximation, the base of $X_7$ must be made large as well, i.e. also the moduli $s^K$, for (at least) a subset $K\in\cK$ with ${\cK \subset \{1,\dots, h^{2,1}_++1\}}$ must be taken to infinity such that the overall volume of $\cB_6$ becomes large, possibly up to shrinkable cycles.\footnote{Even though the gravitational sector in the co-scaled limit is described by supergravity, if $\cK$ is an actual subset of $\{1,\dots,h^{2,1}_++1\}$, there may arise strongly coupled field theory sectors that decouple from the gravitational sector similar to the 4d $\cN=2$ setups studied in~\cite{Marchesano:2023thx,Marchesano:2024tod,Castellano:2024gwi,Castellano:2026bnx}.} Thus, we have to impose the hierarchy 
\begin{equation} \label{eq:hierarchy}
s^K\succsim s^i\,, \qquad K\in \cK\subset \{1,\dots, h^{2,1}_++1\}\,, 
\end{equation} 
to remain within perturbative control. 

In fact, the instability argument for the D$0$-$\overline{{\rm D}0}$ branes can be applied also to the candidate towers of states in other large volume limits in the K\"ahler sector of Type IIA orientifolds. For type II and III limits in the parent $\cN=2$ moduli space, also D$2$-branes wrapping holomorphic curves and D$4$-branes wrapping holomorphic divisors (or bound states thereof) will be part of the leading KK tower in certain limits~\cite{Lee:2019oct}. The argument given here 
carries over, mutatis mutandis, to these  limits. Thus, BPS particles obtained from D2- or D4-branes wrapping the respective holomorphic cycles before the orientifold project to non-BPS, and in particular unstable D$2p$-$\overline{{\rm D}2p}$-bound states, $p\in\{1,2\}$, in the orientifold.
By the same token as above, these states can therefore not serve as the tower of light states predicted by the Distance Conjecture in any classical large volume infinite distance limit in the K\"ahler sector. 

These arguments, and the complementary EFT string perspective of Section \ref{sec:EFTstrings}, show that large K\"ahler limits at constant 4d dilaton and complex structure moduli must be quantum obstructed in Type IIA orientifolds. Furthermore, the quantum corrections not only modify the asymptotic physics, but the infinite distance limits are absent altogether in the quantum moduli space.

\section{Duality to Type IIB Orientifolds} \label{sec:Duality}
 We now compare the absence of the stable KK towers in large volume limits of Type IIA orientifolds to the mirror dual Type IIB/ F-theory setup studied in~\cite{Kaufmann:2026fli,Kaufmann:2026mha}. In Type IIB Calabi--Yau orientifolds with 7-branes, the complex structure moduli of the Calabi--Yau that survive the orientifold projection are the volumes of orientifold odd special Lagrangian 3-cycles~\cite{Grimm:2004ua}. Similar to the situation for D0-branes, only non-BPS and unstable (wrapped) D$3$-$\overline{{\rm D}3}$ particle states exist in the orientifold. As such, these states cannot be the constituents of a leading asymptotically massless tower of states in an infinite distance limit.\footnote{This differs from the conclusions drawn in~\cite{Enriquez-Rojo:2020pqm}.}

  Consider now the orientifold odd 3-cycles with $T^3$ topology which shrink at the fastest rate in the parent $\cN=2$ infinite distance limit~\cite{Hassfeld:2025uoy,Monnee:2025ynn}.
  Even though D3-branes wrapping these cycles do not give rise to massless towers,
  some of these $T^3$s
  can lift to 4-cycles of $T^4$ topology in the F-theory fourfold. In the F-theory lift of the complex structure limit, these $T^4$s shrink to zero size, signalling a residual infinite distance limit in the complex structure moduli space of the fourfold, i.e., in the fully $g_s$-corrected moduli space. As argued in~\cite{Kaufmann:2026mha}, these residual infinite distance limits are generically further obstructed by K\"ahler corrections. However, unlike for $g_s$ corrections, there is no geometric description of the fully $\alpha'$-corrected effective action from which one could infer whether or not the infinite distance regime is completely removed in the quantum moduli space. 

  What comes to the rescue is that the mirror duals of the $g_s$ and K\"ahler corrections, which are naturally distinguished in the Type IIB/F-theory context, are on the same footing in Type IIA orientifolds. Relatedly, O-type A and O-type B limits in Type IIB orientifolds as defined in~\cite{Kaufmann:2026fli} map to the same type of limits in the dual Type IIA orientifold. From the Type IIA perspective, we hence do not expect a physical difference between O-type A and O-type B limits at the quantum level. 
  
  This is particularly striking for type II limits in Type IIB. As shown in~\cite{Kaufmann:2026fli}, $g_s$ corrections completely remove the infinite distance for O-type A limits of type II since there are no $T^4$s in the corresponding F-theory fourfold that shrink in this limit. Instead, type II limits of O-type B trivially lift to F-theory, implying the existence of asymptotically vanishing $T^4$s in the F-theory moduli space. O-type B type~II limits hence give rise to type~II limits also in the F-theory complex structure moduli space. 
  Taking into account also K\"ahler corrections, these limits are obstructed~\cite{Kaufmann:2026mha}, but from the F-theory perspective it is not clear whether the infinite distance behaviour is completely removed at the quantum level or if it is merely not describable within the classical supergravity.
   The dual Type IIA perspective suggests an answer: Here O-type A and O-type B limits are qualitatively the same, and so are the corrections dual to the $g_s$ and K\"ahler moduli corrections. Therefore the effect of the K\"ahler obstructions to type~II  O-type~B limits must be the same as the $g_s$ obstruction to type~II O-type A limits. In other words, the K\"ahler obstructions completely remove the infinite distance limit also for O-type B type II limits. 

This is consistent with our findings in Section~\ref{sec:NoTower}, where we argued that these classical infinite distance limits in Type IIA orientifolds are
   absent
  because there is no infinite tower of light states. For type III and IV limits in Type IIB orientifolds, we do not know of an example in which $g_s$ corrections alone remove the infinite distance limits. However, the arguments of Section \ref{sec:NoTower} did not depend on the type of the singularity. Accordingly, in the dual Type IIB setup also classical infinite distance directions of type III or IV limits of either O-type must be effectively compactified at the quantum level once all $g_s$ and $\alpha'$ corrections are taken into account.
  
  To summarize, the different Type IIA perspectives invoked here draw a consistent and unifying picture also in comparison to their dual Type IIB orientifolds: All infinite distance limits in the moduli space of orientifolds that descend from infinite distance limits in the pure vector multiplet sector of the unorientifolded 4d $\cN=2$ parent theory are removed at the quantum level.

\section{Conclusions}\label{sec:Conclusions}
 In this note, we have extended the investigation of~\cite{Kaufmann:2026fli,Kaufmann:2026mha} of quantum obstructions for classical infinite distance limits in 4d theories of gravity with minimal $\cN=1$ supersymmetry. Whereas the results of~\cite{Kaufmann:2026fli,Kaufmann:2026mha} are based on Type IIB orientifold/F-theory constructions, here we have taken the dual perspective of Type IIA orientifolds/M-theory on $G_2$ manifolds. In this way we have addressed questions left open by the previous analysis. 
  Our focus is on limits which only involve those moduli that descend from vector multiplets of an underlying 4d ${\cal N}=2$ parent theory, i.e. pure complex structure moduli in Type IIB orientifolds or pure K\"ahler moduli limits in Type IIA orientifolds, with the 4d dilaton and all remaining moduli kept finite.
  In the Type IIB orientifold setup, the quantum corrections naturally split into two classes, the $g_s$ corrections discussed in \cite{Kaufmann:2026fli} and K\"ahler corrections, which are the topic of~\cite{Kaufmann:2026mha}. F-theory geometrizes the $g_s$ corrections and therefore provides the right framework to establish that certain classical infinite distance limits are completely removed at the quantum level. For K\"ahler corrections a similar analysis is technically out of reach based on first principles. The central observation underlying this work is that the Type IIA orientifold perspective unifies the dual of $g_s$ and K\"ahler obstructions in a single kind of correction. Moreover, the non-perturbative uplift of Type IIA orientifolds leads to M-theory compactified on a $G_2$ manifold. Compared to the Type IIB uplift given by F-theory on elliptic Calabi--Yau fourfolds, $G_2$ compactifications of M-theory encode even more properties of the low-energy effective action in the geometry. Taking the Type IIA/M-theory perspective, we have been able to make progress on two points that were left open in the Type IIB analysis of~\cite{Kaufmann:2026fli,Kaufmann:2026mha}:
\begin{enumerate}
    \item Using the properties of candidate EFT strings and candidate towers of light states, we have argued that also type IV limits in the Type IIB complex structure moduli space at finite K\"ahler volume are obstructed in the ${\cal N}=1$ context. The regular-fiber type IV limits in the complex structure moduli space of elliptic Calabi--Yau fourfolds are the only limits for which a K\"ahler obstruction could not be established in~\cite{Kaufmann:2026mha}. Our analysis in this note closes this gap. 
    \item We have argued that the quantum obstructions completely remove the classical infinite distance behaviour, thereby effectively compactifying the quantum moduli space in the obstructed direction. The complete removal of an infinite distance limit due to pure $g_s$ corrections was shown for certain type II limits in Type IIB orientifolds using the F-theory uplift in~\cite{Kaufmann:2026fli}. Whether or not K\"ahler corrections can have a similarly drastic effect remained an open question in the analysis of~\cite{Kaufmann:2026mha}. Based on the common origin of the duals of $g_s$ corrections and K\"ahler corrections  in Type IIA orientifolds, we have argued that more generally all classical infinite distance limits that descend from the 4d $\cN=2$ vector multiplet moduli space upon orientifolding are removed at the quantum level. 
\end{enumerate}

We stress that combined limits, i.e.
large complex structure limits in Type IIB with suitably coscaled K\"ahler moduli and their mirror duals in Type IIA orientifolds are, of course, not ruled out by these arguments.
 On the contrary, our results show that only those asymptotic regions can lie in the quantum moduli space in which the remaining moduli are coscaled correspondingly, see \eqref{eq:hierarchy} for the Type IIA/M-theory limits and \cite{Kaufmann:2026mha} for the Type IIB/F-theory duals.

The perspective of Type IIA orientifolds/ M-theory on $G_2$ manifolds hence provides crucial insights into the nature of quantum obstructions in 4d $\cN=1$ effective theories that are not obvious in the Type IIB/F-theory frame studied in~\cite{Kaufmann:2026fli,Kaufmann:2026mha}. Notice, however, that to use the Type IIA perspective we have to focus on the class of classical infinite distance limits that correspond to large volume limits of the underlying Calabi--Yau threefold. Instead, the limits in the Type IIB/F-theory complex structure moduli space that are not connected to the large complex structure point are not covered by our analysis. Nonetheless, given that without orientifolds these limits have very similar properties to the much studied large complex structure loci~\cite{Hassfeld:2025uoy,Monnee:2025ynn}, we expect the qualitative results in this letter to also carry over to these more general limits.  \newline

{\bf Acknowledgements.}
We thank Arthur Hebecker, Dieter L\"ust and Jeroen Monnee for useful discussions. We are particularly grateful to Jeroen Monnee for collaboration on related matters. This work is supported in part by Deutsche Forschungsgemeinschaft under Germany’s Excellence Strategy EXC 2121 Quantum Universe 390833306, by Deutsche Forschungsgemeinschaft through a German-Israeli Project Cooperation (DIP) grant “Holography and the Swampland” and by Deutsche Forschungsgemeinschaft through the Collaborative Research Center 1624 “Higher Structures, Moduli Spaces and Integrability.” MW acknowledges support by  Deutsche Forschungsgemeinschaft through the Emmy Noether program 557478919. 

\appendix

\bibliography{papers_Max}

\begin{thebibliography}{45}%
\makeatletter
\providecommand \@ifxundefined [1]{%
 \@ifx{#1\undefined}
}%
\providecommand \@ifnum [1]{%
 \ifnum #1\expandafter \@firstoftwo
 \else \expandafter \@secondoftwo
 \fi
}%
\providecommand \@ifx [1]{%
 \ifx #1\expandafter \@firstoftwo
 \else \expandafter \@secondoftwo
 \fi
}%
\providecommand \natexlab [1]{#1}%
\providecommand \enquote  [1]{``#1''}%
\providecommand \bibnamefont  [1]{#1}%
\providecommand \bibfnamefont [1]{#1}%
\providecommand \citenamefont [1]{#1}%
\providecommand \href@noop [0]{\@secondoftwo}%
\providecommand \href [0]{\begingroup \@sanitize@url \@href}%
\providecommand \@href[1]{\@@startlink{#1}\@@href}%
\providecommand \@@href[1]{\endgroup#1\@@endlink}%
\providecommand \@sanitize@url [0]{\catcode `\\12\catcode `\$12\catcode
  `\&12\catcode `\#12\catcode `\^12\catcode `\_12\catcode `\%12\relax}%
\providecommand \@@startlink[1]{}%
\providecommand \@@endlink[0]{}%
\providecommand \url  [0]{\begingroup\@sanitize@url \@url }%
\providecommand \@url [1]{\endgroup\@href {#1}{\urlprefix }}%
\providecommand \urlprefix  [0]{URL }%
\providecommand \Eprint [0]{\href }%
\providecommand \doibase [0]{https://doi.org/}%
\providecommand \selectlanguage [0]{\@gobble}%
\providecommand \bibinfo  [0]{\@secondoftwo}%
\providecommand \bibfield  [0]{\@secondoftwo}%
\providecommand \translation [1]{[#1]}%
\providecommand \BibitemOpen [0]{}%
\providecommand \bibitemStop [0]{}%
\providecommand \bibitemNoStop [0]{.\EOS\space}%
\providecommand \EOS [0]{\spacefactor3000\relax}%
\providecommand \BibitemShut  [1]{\csname bibitem#1\endcsname}%
\let\auto@bib@innerbib\@empty
\bibitem [{\citenamefont {Kaufmann}\ \emph
  {et~al.}(2026{\natexlab{a}})\citenamefont {Kaufmann}, \citenamefont {Monnee},
  \citenamefont {Weigand},\ and\ \citenamefont {Wiesner}}]{Kaufmann:2026fli}%
  \BibitemOpen
  \bibfield  {author} {\bibinfo {author} {\bibfnamefont {L.}~\bibnamefont
  {Kaufmann}}, \bibinfo {author} {\bibfnamefont {J.}~\bibnamefont {Monnee}},
  \bibinfo {author} {\bibfnamefont {T.}~\bibnamefont {Weigand}},\ and\ \bibinfo
  {author} {\bibfnamefont {M.}~\bibnamefont {Wiesner}},\ }\bibfield  {title}
  {\bibinfo {title} {{Quantum obstructions for N=1 infinite distance limits --
  Part I: $g_s$ obstructions}},\ }\href@noop {} {\  (\bibinfo {year}
  {2026}{\natexlab{a}})},\ \Eprint {https://arxiv.org/abs/2603.12315}
  {arXiv:2603.12315 [hep-th]} \BibitemShut {NoStop}%
\bibitem [{\citenamefont {Kaufmann}\ \emph
  {et~al.}(2026{\natexlab{b}})\citenamefont {Kaufmann}, \citenamefont {Monnee},
  \citenamefont {Weigand},\ and\ \citenamefont {Wiesner}}]{Kaufmann:2026mha}%
  \BibitemOpen
  \bibfield  {author} {\bibinfo {author} {\bibfnamefont {L.}~\bibnamefont
  {Kaufmann}}, \bibinfo {author} {\bibfnamefont {J.}~\bibnamefont {Monnee}},
  \bibinfo {author} {\bibfnamefont {T.}~\bibnamefont {Weigand}},\ and\ \bibinfo
  {author} {\bibfnamefont {M.}~\bibnamefont {Wiesner}},\ }\bibfield  {title}
  {\bibinfo {title} {{Quantum obstructions for N=1 infinite distance limits --
  Part II: K{\"a}hler obstructions}},\ }\href@noop {} {\  (\bibinfo {year}
  {2026}{\natexlab{b}})},\ \Eprint {https://arxiv.org/abs/2603.13470}
  {arXiv:2603.13470 [hep-th]} \BibitemShut {NoStop}%
\bibitem [{\citenamefont {Lanza}\ \emph {et~al.}(2021)\citenamefont {Lanza},
  \citenamefont {Marchesano}, \citenamefont {Martucci},\ and\ \citenamefont
  {Valenzuela}}]{Lanza:2021udy}%
  \BibitemOpen
  \bibfield  {author} {\bibinfo {author} {\bibfnamefont {S.}~\bibnamefont
  {Lanza}}, \bibinfo {author} {\bibfnamefont {F.}~\bibnamefont {Marchesano}},
  \bibinfo {author} {\bibfnamefont {L.}~\bibnamefont {Martucci}},\ and\
  \bibinfo {author} {\bibfnamefont {I.}~\bibnamefont {Valenzuela}},\ }\bibfield
   {title} {\bibinfo {title} {{The EFT stringy viewpoint on large distances}},\
  }\href {https://doi.org/10.1007/JHEP09(2021)197} {\bibfield  {journal}
  {\bibinfo  {journal} {JHEP}\ }\textbf {\bibinfo {volume} {09}},\ \bibinfo
  {pages} {197}},\ \Eprint {https://arxiv.org/abs/2104.05726} {arXiv:2104.05726
  [hep-th]} \BibitemShut {NoStop}%
\bibitem [{\citenamefont {Martucci}\ \emph {et~al.}(2023)\citenamefont
  {Martucci}, \citenamefont {Risso},\ and\ \citenamefont
  {Weigand}}]{Martucci:2022krl}%
  \BibitemOpen
  \bibfield  {author} {\bibinfo {author} {\bibfnamefont {L.}~\bibnamefont
  {Martucci}}, \bibinfo {author} {\bibfnamefont {N.}~\bibnamefont {Risso}},\
  and\ \bibinfo {author} {\bibfnamefont {T.}~\bibnamefont {Weigand}},\
  }\bibfield  {title} {\bibinfo {title} {{Quantum gravity bounds on $
  \mathcal{N} $ = 1 effective theories in four dimensions}},\ }\href
  {https://doi.org/10.1007/JHEP03(2023)197} {\bibfield  {journal} {\bibinfo
  {journal} {JHEP}\ }\textbf {\bibinfo {volume} {03}},\ \bibinfo {pages}
  {197}},\ \Eprint {https://arxiv.org/abs/2210.10797} {arXiv:2210.10797
  [hep-th]} \BibitemShut {NoStop}%
\bibitem [{\citenamefont {Marchesano}\ and\ \citenamefont
  {Melotti}(2023)}]{Marchesano:2022axe}%
  \BibitemOpen
  \bibfield  {author} {\bibinfo {author} {\bibfnamefont {F.}~\bibnamefont
  {Marchesano}}\ and\ \bibinfo {author} {\bibfnamefont {L.}~\bibnamefont
  {Melotti}},\ }\bibfield  {title} {\bibinfo {title} {{EFT strings and
  emergence}},\ }\href {https://doi.org/10.1007/JHEP02(2023)112} {\bibfield
  {journal} {\bibinfo  {journal} {JHEP}\ }\textbf {\bibinfo {volume} {02}},\
  \bibinfo {pages} {112}},\ \Eprint {https://arxiv.org/abs/2211.01409}
  {arXiv:2211.01409 [hep-th]} \BibitemShut {NoStop}%
\bibitem [{\citenamefont {Martucci}\ \emph {et~al.}(2024)\citenamefont
  {Martucci}, \citenamefont {Risso}, \citenamefont {Valenti},\ and\
  \citenamefont {Vecchi}}]{Martucci:2024trp}%
  \BibitemOpen
  \bibfield  {author} {\bibinfo {author} {\bibfnamefont {L.}~\bibnamefont
  {Martucci}}, \bibinfo {author} {\bibfnamefont {N.}~\bibnamefont {Risso}},
  \bibinfo {author} {\bibfnamefont {A.}~\bibnamefont {Valenti}},\ and\ \bibinfo
  {author} {\bibfnamefont {L.}~\bibnamefont {Vecchi}},\ }\bibfield  {title}
  {\bibinfo {title} {{Wormholes in the axiverse, and the species scale}},\
  }\href {https://doi.org/10.1007/JHEP07(2024)240} {\bibfield  {journal}
  {\bibinfo  {journal} {JHEP}\ }\textbf {\bibinfo {volume} {07}},\ \bibinfo
  {pages} {240}},\ \Eprint {https://arxiv.org/abs/2404.14489} {arXiv:2404.14489
  [hep-th]} \BibitemShut {NoStop}%
\bibitem [{\citenamefont {Grieco}\ \emph {et~al.}(2025)\citenamefont {Grieco},
  \citenamefont {Ruiz},\ and\ \citenamefont {Valenzuela}}]{Grieco:2025bjy}%
  \BibitemOpen
  \bibfield  {author} {\bibinfo {author} {\bibfnamefont {A.}~\bibnamefont
  {Grieco}}, \bibinfo {author} {\bibfnamefont {I.}~\bibnamefont {Ruiz}},\ and\
  \bibinfo {author} {\bibfnamefont {I.}~\bibnamefont {Valenzuela}},\ }\bibfield
   {title} {\bibinfo {title} {{EFT strings and dualities in 4d
  $\mathcal{N}=1$}},\ }\href@noop {} {\  (\bibinfo {year} {2025})},\ \Eprint
  {https://arxiv.org/abs/2504.16984} {arXiv:2504.16984 [hep-th]} \BibitemShut
  {NoStop}%
\bibitem [{\citenamefont {Hassfeld}\ \emph {et~al.}(2026)\citenamefont
  {Hassfeld}, \citenamefont {Monnee}, \citenamefont {Weigand},\ and\
  \citenamefont {Wiesner}}]{Hassfeld:2025uoy}%
  \BibitemOpen
  \bibfield  {author} {\bibinfo {author} {\bibfnamefont {B.}~\bibnamefont
  {Hassfeld}}, \bibinfo {author} {\bibfnamefont {J.}~\bibnamefont {Monnee}},
  \bibinfo {author} {\bibfnamefont {T.}~\bibnamefont {Weigand}},\ and\ \bibinfo
  {author} {\bibfnamefont {M.}~\bibnamefont {Wiesner}},\ }\bibfield  {title}
  {\bibinfo {title} {{Emergent strings in Type IIB Calabi-Yau
  compactifications}},\ }\href {https://doi.org/10.1007/JHEP01(2026)140}
  {\bibfield  {journal} {\bibinfo  {journal} {JHEP}\ }\textbf {\bibinfo
  {volume} {01}},\ \bibinfo {pages} {140}},\ \Eprint
  {https://arxiv.org/abs/2504.01066} {arXiv:2504.01066 [hep-th]} \BibitemShut
  {NoStop}%
\bibitem [{\citenamefont {Monnee}\ \emph {et~al.}(2026)\citenamefont {Monnee},
  \citenamefont {Weigand},\ and\ \citenamefont {Wiesner}}]{Monnee:2025ynn}%
  \BibitemOpen
  \bibfield  {author} {\bibinfo {author} {\bibfnamefont {J.}~\bibnamefont
  {Monnee}}, \bibinfo {author} {\bibfnamefont {T.}~\bibnamefont {Weigand}},\
  and\ \bibinfo {author} {\bibfnamefont {M.}~\bibnamefont {Wiesner}},\
  }\bibfield  {title} {\bibinfo {title} {{Physics and geometry of complex
  structure limits in type IIB Calabi-Yau compactifications}},\ }\href
  {https://doi.org/10.1007/JHEP03(2026)063} {\bibfield  {journal} {\bibinfo
  {journal} {JHEP}\ }\textbf {\bibinfo {volume} {03}},\ \bibinfo {pages}
  {063}},\ \Eprint {https://arxiv.org/abs/2509.07056} {arXiv:2509.07056
  [hep-th]} \BibitemShut {NoStop}%
\bibitem [{\citenamefont {Grimm}\ \emph {et~al.}(2018)\citenamefont {Grimm},
  \citenamefont {Palti},\ and\ \citenamefont {Valenzuela}}]{Grimm:2018ohb}%
  \BibitemOpen
  \bibfield  {author} {\bibinfo {author} {\bibfnamefont {T.~W.}\ \bibnamefont
  {Grimm}}, \bibinfo {author} {\bibfnamefont {E.}~\bibnamefont {Palti}},\ and\
  \bibinfo {author} {\bibfnamefont {I.}~\bibnamefont {Valenzuela}},\ }\bibfield
   {title} {\bibinfo {title} {{Infinite Distances in Field Space and Massless
  Towers of States}},\ }\href {https://doi.org/10.1007/JHEP08(2018)143}
  {\bibfield  {journal} {\bibinfo  {journal} {JHEP}\ }\textbf {\bibinfo
  {volume} {08}},\ \bibinfo {pages} {143}},\ \Eprint
  {https://arxiv.org/abs/1802.08264} {arXiv:1802.08264 [hep-th]} \BibitemShut
  {NoStop}%
\bibitem [{\citenamefont {Grimm}\ \emph {et~al.}(2019)\citenamefont {Grimm},
  \citenamefont {Li},\ and\ \citenamefont {Palti}}]{Grimm:2018cpv}%
  \BibitemOpen
  \bibfield  {author} {\bibinfo {author} {\bibfnamefont {T.~W.}\ \bibnamefont
  {Grimm}}, \bibinfo {author} {\bibfnamefont {C.}~\bibnamefont {Li}},\ and\
  \bibinfo {author} {\bibfnamefont {E.}~\bibnamefont {Palti}},\ }\bibfield
  {title} {\bibinfo {title} {{Infinite Distance Networks in Field Space and
  Charge Orbits}},\ }\href {https://doi.org/10.1007/JHEP03(2019)016} {\bibfield
   {journal} {\bibinfo  {journal} {JHEP}\ }\textbf {\bibinfo {volume} {03}},\
  \bibinfo {pages} {016}},\ \Eprint {https://arxiv.org/abs/1811.02571}
  {arXiv:1811.02571 [hep-th]} \BibitemShut {NoStop}%
\bibitem [{\citenamefont {Corvilain}\ \emph {et~al.}(2019)\citenamefont
  {Corvilain}, \citenamefont {Grimm},\ and\ \citenamefont
  {Valenzuela}}]{Corvilain:2018lgw}%
  \BibitemOpen
  \bibfield  {author} {\bibinfo {author} {\bibfnamefont {P.}~\bibnamefont
  {Corvilain}}, \bibinfo {author} {\bibfnamefont {T.~W.}\ \bibnamefont
  {Grimm}},\ and\ \bibinfo {author} {\bibfnamefont {I.}~\bibnamefont
  {Valenzuela}},\ }\bibfield  {title} {\bibinfo {title} {{The Swampland
  Distance Conjecture for Kahler moduli}},\ }\href
  {https://doi.org/10.1007/JHEP08(2019)075} {\bibfield  {journal} {\bibinfo
  {journal} {JHEP}\ }\textbf {\bibinfo {volume} {08}},\ \bibinfo {pages}
  {075}},\ \Eprint {https://arxiv.org/abs/1812.07548} {arXiv:1812.07548
  [hep-th]} \BibitemShut {NoStop}%
\bibitem [{\citenamefont {Lee}\ \emph {et~al.}(2022)\citenamefont {Lee},
  \citenamefont {Lerche},\ and\ \citenamefont {Weigand}}]{Lee:2019oct}%
  \BibitemOpen
  \bibfield  {author} {\bibinfo {author} {\bibfnamefont {S.-J.}\ \bibnamefont
  {Lee}}, \bibinfo {author} {\bibfnamefont {W.}~\bibnamefont {Lerche}},\ and\
  \bibinfo {author} {\bibfnamefont {T.}~\bibnamefont {Weigand}},\ }\bibfield
  {title} {\bibinfo {title} {{Emergent strings from infinite distance
  limits}},\ }\href {https://doi.org/10.1007/JHEP02(2022)190} {\bibfield
  {journal} {\bibinfo  {journal} {JHEP}\ }\textbf {\bibinfo {volume} {02}},\
  \bibinfo {pages} {190}},\ \Eprint {https://arxiv.org/abs/1910.01135}
  {arXiv:1910.01135 [hep-th]} \BibitemShut {NoStop}%
\bibitem [{\citenamefont {Katz}\ \emph {et~al.}(2020)\citenamefont {Katz},
  \citenamefont {Kim}, \citenamefont {Tarazi},\ and\ \citenamefont
  {Vafa}}]{Katz:2020ewz}%
  \BibitemOpen
  \bibfield  {author} {\bibinfo {author} {\bibfnamefont {S.}~\bibnamefont
  {Katz}}, \bibinfo {author} {\bibfnamefont {H.-C.}\ \bibnamefont {Kim}},
  \bibinfo {author} {\bibfnamefont {H.-C.}\ \bibnamefont {Tarazi}},\ and\
  \bibinfo {author} {\bibfnamefont {C.}~\bibnamefont {Vafa}},\ }\bibfield
  {title} {\bibinfo {title} {{Swampland Constraints on 5d $\mathcal{N}=1$
  Supergravity}},\ }\href {https://doi.org/10.1007/JHEP07(2020)080} {\bibfield
  {journal} {\bibinfo  {journal} {JHEP}\ }\textbf {\bibinfo {volume} {07}},\
  \bibinfo {pages} {080}},\ \Eprint {https://arxiv.org/abs/2004.14401}
  {arXiv:2004.14401 [hep-th]} \BibitemShut {NoStop}%
\bibitem [{\citenamefont {Kim}\ \emph {et~al.}(2019)\citenamefont {Kim},
  \citenamefont {Shiu},\ and\ \citenamefont {Vafa}}]{Kim:2019vuc}%
  \BibitemOpen
  \bibfield  {author} {\bibinfo {author} {\bibfnamefont {H.-C.}\ \bibnamefont
  {Kim}}, \bibinfo {author} {\bibfnamefont {G.}~\bibnamefont {Shiu}},\ and\
  \bibinfo {author} {\bibfnamefont {C.}~\bibnamefont {Vafa}},\ }\bibfield
  {title} {\bibinfo {title} {{Branes and the Swampland}},\ }\href
  {https://doi.org/10.1103/PhysRevD.100.066006} {\bibfield  {journal} {\bibinfo
   {journal} {Phys. Rev. D}\ }\textbf {\bibinfo {volume} {100}},\ \bibinfo
  {pages} {066006} (\bibinfo {year} {2019})},\ \Eprint
  {https://arxiv.org/abs/1905.08261} {arXiv:1905.08261 [hep-th]} \BibitemShut
  {NoStop}%
\bibitem [{\citenamefont {Ooguri}\ and\ \citenamefont
  {Vafa}(2007)}]{Ooguri:2006in}%
  \BibitemOpen
  \bibfield  {author} {\bibinfo {author} {\bibfnamefont {H.}~\bibnamefont
  {Ooguri}}\ and\ \bibinfo {author} {\bibfnamefont {C.}~\bibnamefont {Vafa}},\
  }\bibfield  {title} {\bibinfo {title} {{On the Geometry of the String
  Landscape and the Swampland}},\ }\href
  {https://doi.org/10.1016/j.nuclphysb.2006.10.033} {\bibfield  {journal}
  {\bibinfo  {journal} {Nucl. Phys.}\ }\textbf {\bibinfo {volume} {B766}},\
  \bibinfo {pages} {21} (\bibinfo {year} {2007})},\ \Eprint
  {https://arxiv.org/abs/hep-th/0605264} {arXiv:hep-th/0605264 [hep-th]}
  \BibitemShut {NoStop}%
\bibitem [{\citenamefont {Grimm}\ and\ \citenamefont
  {Louis}(2005)}]{Grimm:2004ua}%
  \BibitemOpen
  \bibfield  {author} {\bibinfo {author} {\bibfnamefont {T.~W.}\ \bibnamefont
  {Grimm}}\ and\ \bibinfo {author} {\bibfnamefont {J.}~\bibnamefont {Louis}},\
  }\bibfield  {title} {\bibinfo {title} {{The Effective action of type IIA
  Calabi-Yau orientifolds}},\ }\href
  {https://doi.org/10.1016/j.nuclphysb.2005.04.007} {\bibfield  {journal}
  {\bibinfo  {journal} {Nucl. Phys. B}\ }\textbf {\bibinfo {volume} {718}},\
  \bibinfo {pages} {153} (\bibinfo {year} {2005})},\ \Eprint
  {https://arxiv.org/abs/hep-th/0412277} {arXiv:hep-th/0412277} \BibitemShut
  {NoStop}%
\bibitem [{\citenamefont {Grimm}(2005)}]{Grimm:2005fa}%
  \BibitemOpen
  \bibfield  {author} {\bibinfo {author} {\bibfnamefont {T.~W.}\ \bibnamefont
  {Grimm}},\ }\bibfield  {title} {\bibinfo {title} {{The Effective action of
  type II Calabi-Yau orientifolds}},\ }\href
  {https://doi.org/10.1002/prop.200510253} {\bibfield  {journal} {\bibinfo
  {journal} {Fortsch. Phys.}\ }\textbf {\bibinfo {volume} {53}},\ \bibinfo
  {pages} {1179} (\bibinfo {year} {2005})},\ \Eprint
  {https://arxiv.org/abs/hep-th/0507153} {arXiv:hep-th/0507153 [hep-th]}
  \BibitemShut {NoStop}%
\bibitem [{\citenamefont {Kachru}\ and\ \citenamefont
  {McGreevy}(2001)}]{Kachru:2001je}%
  \BibitemOpen
  \bibfield  {author} {\bibinfo {author} {\bibfnamefont {S.}~\bibnamefont
  {Kachru}}\ and\ \bibinfo {author} {\bibfnamefont {J.}~\bibnamefont
  {McGreevy}},\ }\bibfield  {title} {\bibinfo {title} {{M theory on manifolds
  of G(2) holonomy and type IIA orientifolds}},\ }\href
  {https://doi.org/10.1088/1126-6708/2001/06/027} {\bibfield  {journal}
  {\bibinfo  {journal} {JHEP}\ }\textbf {\bibinfo {volume} {06}},\ \bibinfo
  {pages} {027}},\ \Eprint {https://arxiv.org/abs/hep-th/0103223}
  {arXiv:hep-th/0103223} \BibitemShut {NoStop}%
\bibitem [{\citenamefont {Townsend}(1995)}]{Townsend:1995kk}%
  \BibitemOpen
  \bibfield  {author} {\bibinfo {author} {\bibfnamefont {P.~K.}\ \bibnamefont
  {Townsend}},\ }\bibfield  {title} {\bibinfo {title} {{The eleven-dimensional
  supermembrane revisited}},\ }\href
  {https://doi.org/10.1016/0370-2693(95)00397-4} {\bibfield  {journal}
  {\bibinfo  {journal} {Phys. Lett. B}\ }\textbf {\bibinfo {volume} {350}},\
  \bibinfo {pages} {184} (\bibinfo {year} {1995})},\ \Eprint
  {https://arxiv.org/abs/hep-th/9501068} {arXiv:hep-th/9501068} \BibitemShut
  {NoStop}%
\bibitem [{\citenamefont {Atiyah}\ and\ \citenamefont
  {Hitchin}(1988)}]{Atiyah:1988jp}%
  \BibitemOpen
  \bibfield  {author} {\bibinfo {author} {\bibfnamefont {M.~F.}\ \bibnamefont
  {Atiyah}}\ and\ \bibinfo {author} {\bibfnamefont {N.~J.}\ \bibnamefont
  {Hitchin}},\ }\href@noop {} {\emph {\bibinfo {title} {{The geometry and
  dynamics of magnetic monopoles. M.B. Porter Lectures}}}}\ (\bibinfo {year}
  {1988})\BibitemShut {NoStop}%
\bibitem [{\citenamefont {Seiberg}\ and\ \citenamefont
  {Witten}(1996)}]{Seiberg:1996nz}%
  \BibitemOpen
  \bibfield  {author} {\bibinfo {author} {\bibfnamefont {N.}~\bibnamefont
  {Seiberg}}\ and\ \bibinfo {author} {\bibfnamefont {E.}~\bibnamefont
  {Witten}},\ }\bibfield  {title} {\bibinfo {title} {{Gauge dynamics and
  compactification to three-dimensions}},\ }in\ \href@noop {} {\emph {\bibinfo
  {booktitle} {{Conference on the Mathematical Beauty of Physics (In Memory of
  C. Itzykson)}}}}\ (\bibinfo {year} {1996})\ pp.\ \bibinfo {pages}
  {333--366},\ \Eprint {https://arxiv.org/abs/hep-th/9607163}
  {arXiv:hep-th/9607163} \BibitemShut {NoStop}%
\bibitem [{\citenamefont {Joyce}(2000)}]{10.1093/oso/9780198506010.001.0001}%
  \BibitemOpen
  \bibfield  {author} {\bibinfo {author} {\bibfnamefont {D.~D.}\ \bibnamefont
  {Joyce}},\ }\href {https://doi.org/10.1093/oso/9780198506010.001.0001} {\emph
  {\bibinfo {title} {Compact Manifolds with Special Holonomy}}}\ (\bibinfo
  {publisher} {Oxford University Press},\ \bibinfo {year} {2000})\BibitemShut
  {NoStop}%
\bibitem [{\citenamefont {Acharya}(2000)}]{Acharya:2000gb}%
  \BibitemOpen
  \bibfield  {author} {\bibinfo {author} {\bibfnamefont {B.~S.}\ \bibnamefont
  {Acharya}},\ }\bibfield  {title} {\bibinfo {title} {{On Realizing N=1
  superYang-Mills in M theory}},\ }\href@noop {} {\  (\bibinfo {year}
  {2000})},\ \Eprint {https://arxiv.org/abs/hep-th/0011089}
  {arXiv:hep-th/0011089} \BibitemShut {NoStop}%
\bibitem [{\citenamefont {Acharya}\ and\ \citenamefont
  {Witten}(2001)}]{Acharya:2001gy}%
  \BibitemOpen
  \bibfield  {author} {\bibinfo {author} {\bibfnamefont {B.~S.}\ \bibnamefont
  {Acharya}}\ and\ \bibinfo {author} {\bibfnamefont {E.}~\bibnamefont
  {Witten}},\ }\bibfield  {title} {\bibinfo {title} {{Chiral fermions from
  manifolds of G(2) holonomy}},\ }\href@noop {} {\  (\bibinfo {year} {2001})},\
  \Eprint {https://arxiv.org/abs/hep-th/0109152} {arXiv:hep-th/0109152}
  \BibitemShut {NoStop}%
\bibitem [{\citenamefont {Braun}(2020)}]{Braun:2019wnj}%
  \BibitemOpen
  \bibfield  {author} {\bibinfo {author} {\bibfnamefont {A.~P.}\ \bibnamefont
  {Braun}},\ }\bibfield  {title} {\bibinfo {title} {{M-Theory and
  Orientifolds}},\ }\href {https://doi.org/10.1007/JHEP09(2020)065} {\bibfield
  {journal} {\bibinfo  {journal} {JHEP}\ }\textbf {\bibinfo {volume} {09}},\
  \bibinfo {pages} {065}},\ \Eprint {https://arxiv.org/abs/1912.06072}
  {arXiv:1912.06072 [hep-th]} \BibitemShut {NoStop}%
\bibitem [{\citenamefont {Cota}\ \emph {et~al.}(2022)\citenamefont {Cota},
  \citenamefont {Mininno}, \citenamefont {Weigand},\ and\ \citenamefont
  {Wiesner}}]{Cota:2022yjw}%
  \BibitemOpen
  \bibfield  {author} {\bibinfo {author} {\bibfnamefont {C.~F.}\ \bibnamefont
  {Cota}}, \bibinfo {author} {\bibfnamefont {A.}~\bibnamefont {Mininno}},
  \bibinfo {author} {\bibfnamefont {T.}~\bibnamefont {Weigand}},\ and\ \bibinfo
  {author} {\bibfnamefont {M.}~\bibnamefont {Wiesner}},\ }\bibfield  {title}
  {\bibinfo {title} {{The asymptotic Weak Gravity Conjecture for open
  strings}},\ }\href {https://doi.org/10.1007/JHEP11(2022)058} {\bibfield
  {journal} {\bibinfo  {journal} {JHEP}\ }\textbf {\bibinfo {volume} {11}},\
  \bibinfo {pages} {058}},\ \Eprint {https://arxiv.org/abs/2208.00009}
  {arXiv:2208.00009 [hep-th]} \BibitemShut {NoStop}%
\bibitem [{\citenamefont {Coleman}(1973)}]{Coleman:1973ci}%
  \BibitemOpen
  \bibfield  {author} {\bibinfo {author} {\bibfnamefont {S.~R.}\ \bibnamefont
  {Coleman}},\ }\bibfield  {title} {\bibinfo {title} {{There are no Goldstone
  bosons in two-dimensions}},\ }\href {https://doi.org/10.1007/BF01646487}
  {\bibfield  {journal} {\bibinfo  {journal} {Commun. Math. Phys.}\ }\textbf
  {\bibinfo {volume} {31}},\ \bibinfo {pages} {259} (\bibinfo {year}
  {1973})}\BibitemShut {NoStop}%
\bibitem [{\citenamefont {Mclean}(1998)}]{Mclean}%
  \BibitemOpen
  \bibfield  {author} {\bibinfo {author} {\bibfnamefont {R.~C.}\ \bibnamefont
  {Mclean}},\ }\bibfield  {title} {\bibinfo {title} {Deformations of calibrated
  submanifolds},\ }\href {https://api.semanticscholar.org/CorpusID:52951870}
  {\bibfield  {journal} {\bibinfo  {journal} {Communications in Analysis and
  Geometry}\ }\textbf {\bibinfo {volume} {6}},\ \bibinfo {pages} {705}
  (\bibinfo {year} {1998})}\BibitemShut {NoStop}%
\bibitem [{\citenamefont {Maldacena}\ \emph {et~al.}(1997)\citenamefont
  {Maldacena}, \citenamefont {Strominger},\ and\ \citenamefont
  {Witten}}]{Maldacena:1997de}%
  \BibitemOpen
  \bibfield  {author} {\bibinfo {author} {\bibfnamefont {J.~M.}\ \bibnamefont
  {Maldacena}}, \bibinfo {author} {\bibfnamefont {A.}~\bibnamefont
  {Strominger}},\ and\ \bibinfo {author} {\bibfnamefont {E.}~\bibnamefont
  {Witten}},\ }\bibfield  {title} {\bibinfo {title} {{Black hole entropy in M
  theory}},\ }\href {https://doi.org/10.1088/1126-6708/1997/12/002} {\bibfield
  {journal} {\bibinfo  {journal} {JHEP}\ }\textbf {\bibinfo {volume} {12}},\
  \bibinfo {pages} {002}},\ \Eprint {https://arxiv.org/abs/hep-th/9711053}
  {arXiv:hep-th/9711053} \BibitemShut {NoStop}%
\bibitem [{\citenamefont {Oguiso}(1993)}]{Oguiso}%
  \BibitemOpen
  \bibfield  {author} {\bibinfo {author} {\bibfnamefont {K.}~\bibnamefont
  {Oguiso}},\ }\bibfield  {title} {\bibinfo {title} {{On algebraic fiber space
  structures on a Calabi-Yau 3-fold}},\ }\href
  {https://doi.org/10.1142/S0129167X93000248} {\bibfield  {journal} {\bibinfo
  {journal} {International Journal of Mathematics}\ }\textbf {\bibinfo {volume}
  {04}},\ \bibinfo {pages} {439} (\bibinfo {year} {1993})},\ \Eprint
  {https://arxiv.org/abs/https://doi.org/10.1142/S0129167X93000248}
  {https://doi.org/10.1142/S0129167X93000248} \BibitemShut {NoStop}%
\bibitem [{\citenamefont {Iitaka}(1982)}]{Iitaka}%
  \BibitemOpen
  \bibfield  {author} {\bibinfo {author} {\bibfnamefont {S.}~\bibnamefont
  {Iitaka}},\ }\href@noop {} {\emph {\bibinfo {title} {Algebraic geometry. {A}n
  introduction to birational geometry of algebraic varieties}}},\ \bibinfo
  {series} {Graduate Texts in Mathematics}, Vol.~\bibinfo {volume} {76}\
  (\bibinfo  {publisher} {Springer-Verlag},\ \bibinfo {year}
  {1982})\BibitemShut {NoStop}%
\bibitem [{\citenamefont {Lazarsfeld}(2004)}]{lazarsfeld}%
  \BibitemOpen
  \bibfield  {author} {\bibinfo {author} {\bibfnamefont {R.}~\bibnamefont
  {Lazarsfeld}},\ }\href {https://books.google.de/books?id=jAWVmIz80A4C} {\emph
  {\bibinfo {title} {Positivity in Algebraic Geometry I: Classical Setting:
  Line Bundles and Linear Series}}},\ Ergebnisse der Mathematik und ihrer
  Grenzgebiete. 3. Folge A Series of Modern Surveys in Mathematics\ (\bibinfo
  {publisher} {Springer},\ \bibinfo {year} {2004})\BibitemShut {NoStop}%
\bibitem [{\citenamefont {van Beest}\ \emph {et~al.}(2022)\citenamefont {van
  Beest}, \citenamefont {Calder\'on-Infante}, \citenamefont {Mirfendereski},\
  and\ \citenamefont {Valenzuela}}]{vanBeest:2021lhn}%
  \BibitemOpen
  \bibfield  {author} {\bibinfo {author} {\bibfnamefont {M.}~\bibnamefont {van
  Beest}}, \bibinfo {author} {\bibfnamefont {J.}~\bibnamefont
  {Calder\'on-Infante}}, \bibinfo {author} {\bibfnamefont {D.}~\bibnamefont
  {Mirfendereski}},\ and\ \bibinfo {author} {\bibfnamefont {I.}~\bibnamefont
  {Valenzuela}},\ }\bibfield  {title} {\bibinfo {title} {{Lectures on the
  Swampland Program in String Compactifications}},\ }\href
  {https://doi.org/10.1016/j.physrep.2022.09.002} {\bibfield  {journal}
  {\bibinfo  {journal} {Phys. Rept.}\ }\textbf {\bibinfo {volume} {989}},\
  \bibinfo {pages} {1} (\bibinfo {year} {2022})},\ \Eprint
  {https://arxiv.org/abs/2102.01111} {arXiv:2102.01111 [hep-th]} \BibitemShut
  {NoStop}%
\bibitem [{\citenamefont {Witten}(1996)}]{Witten:1996mz}%
  \BibitemOpen
  \bibfield  {author} {\bibinfo {author} {\bibfnamefont {E.}~\bibnamefont
  {Witten}},\ }\bibfield  {title} {\bibinfo {title} {{Strong coupling expansion
  of Calabi-Yau compactification}},\ }\href
  {https://doi.org/10.1016/0550-3213(96)00190-3} {\bibfield  {journal}
  {\bibinfo  {journal} {Nucl. Phys. B}\ }\textbf {\bibinfo {volume} {471}},\
  \bibinfo {pages} {135} (\bibinfo {year} {1996})},\ \Eprint
  {https://arxiv.org/abs/hep-th/9602070} {arXiv:hep-th/9602070} \BibitemShut
  {NoStop}%
\bibitem [{\citenamefont {Cveti{\v{c}}}\ and\ \citenamefont
  {Wiesner}(2024)}]{Cvetic:2024wsj}%
  \BibitemOpen
  \bibfield  {author} {\bibinfo {author} {\bibfnamefont {M.}~\bibnamefont
  {Cveti{\v{c}}}}\ and\ \bibinfo {author} {\bibfnamefont {M.}~\bibnamefont
  {Wiesner}},\ }\bibfield  {title} {\bibinfo {title} {{Nonperturbative
  resolution of strong coupling singularities in 4D N=1 heterotic M-theory}},\
  }\href {https://doi.org/10.1103/PhysRevD.110.106008} {\bibfield  {journal}
  {\bibinfo  {journal} {Phys. Rev. D}\ }\textbf {\bibinfo {volume} {110}},\
  \bibinfo {pages} {106008} (\bibinfo {year} {2024})},\ \Eprint
  {https://arxiv.org/abs/2408.12458} {arXiv:2408.12458 [hep-th]} \BibitemShut
  {NoStop}%
\bibitem [{\citenamefont {Cveti{\v{c}}}\ and\ \citenamefont
  {Wiesner}(2026)}]{Cvetic:2025nfx}%
  \BibitemOpen
  \bibfield  {author} {\bibinfo {author} {\bibfnamefont {M.}~\bibnamefont
  {Cveti{\v{c}}}}\ and\ \bibinfo {author} {\bibfnamefont {M.}~\bibnamefont
  {Wiesner}},\ }\bibfield  {title} {\bibinfo {title} {{Smooth String Vacua in a
  Gravitationally Nonperturbative Regime}},\ }\href
  {https://doi.org/10.1103/cty8-hg73} {\bibfield  {journal} {\bibinfo
  {journal} {Phys. Rev. Lett.}\ }\textbf {\bibinfo {volume} {136}},\ \bibinfo
  {pages} {121602} (\bibinfo {year} {2026})},\ \Eprint
  {https://arxiv.org/abs/2511.01975} {arXiv:2511.01975 [hep-th]} \BibitemShut
  {NoStop}%
\bibitem [{\citenamefont {Lust}\ and\ \citenamefont
  {Stieberger}(2007)}]{Lust:2003ky}%
  \BibitemOpen
  \bibfield  {author} {\bibinfo {author} {\bibfnamefont {D.}~\bibnamefont
  {Lust}}\ and\ \bibinfo {author} {\bibfnamefont {S.}~\bibnamefont
  {Stieberger}},\ }\bibfield  {title} {\bibinfo {title} {{Gauge threshold
  corrections in intersecting brane world models}},\ }\href
  {https://doi.org/10.1002/prop.200310335} {\bibfield  {journal} {\bibinfo
  {journal} {Fortsch. Phys.}\ }\textbf {\bibinfo {volume} {55}},\ \bibinfo
  {pages} {427} (\bibinfo {year} {2007})},\ \Eprint
  {https://arxiv.org/abs/hep-th/0302221} {arXiv:hep-th/0302221} \BibitemShut
  {NoStop}%
\bibitem [{\citenamefont {Akerblom}\ \emph {et~al.}(2007)\citenamefont
  {Akerblom}, \citenamefont {Blumenhagen}, \citenamefont {Lust},\ and\
  \citenamefont {Schmidt-Sommerfeld}}]{Akerblom:2007np}%
  \BibitemOpen
  \bibfield  {author} {\bibinfo {author} {\bibfnamefont {N.}~\bibnamefont
  {Akerblom}}, \bibinfo {author} {\bibfnamefont {R.}~\bibnamefont
  {Blumenhagen}}, \bibinfo {author} {\bibfnamefont {D.}~\bibnamefont {Lust}},\
  and\ \bibinfo {author} {\bibfnamefont {M.}~\bibnamefont
  {Schmidt-Sommerfeld}},\ }\bibfield  {title} {\bibinfo {title} {{Thresholds
  for Intersecting D-branes Revisited}},\ }\href
  {https://doi.org/10.1016/j.physletb.2007.06.060} {\bibfield  {journal}
  {\bibinfo  {journal} {Phys. Lett. B}\ }\textbf {\bibinfo {volume} {652}},\
  \bibinfo {pages} {53} (\bibinfo {year} {2007})},\ \Eprint
  {https://arxiv.org/abs/0705.2150} {arXiv:0705.2150 [hep-th]} \BibitemShut
  {NoStop}%
\bibitem [{\citenamefont {Blumenhagen}\ and\ \citenamefont
  {Schmidt-Sommerfeld}(2007)}]{Blumenhagen:2007ip}%
  \BibitemOpen
  \bibfield  {author} {\bibinfo {author} {\bibfnamefont {R.}~\bibnamefont
  {Blumenhagen}}\ and\ \bibinfo {author} {\bibfnamefont {M.}~\bibnamefont
  {Schmidt-Sommerfeld}},\ }\bibfield  {title} {\bibinfo {title} {{Gauge
  Thresholds and Kaehler Metrics for Rigid Intersecting D-brane Models}},\
  }\href {https://doi.org/10.1088/1126-6708/2007/12/072} {\bibfield  {journal}
  {\bibinfo  {journal} {JHEP}\ }\textbf {\bibinfo {volume} {12}},\ \bibinfo
  {pages} {072}},\ \Eprint {https://arxiv.org/abs/0711.0866} {arXiv:0711.0866
  [hep-th]} \BibitemShut {NoStop}%
\bibitem [{\citenamefont {Marchesano}\ \emph {et~al.}(2024)\citenamefont
  {Marchesano}, \citenamefont {Melotti},\ and\ \citenamefont
  {Paoloni}}]{Marchesano:2023thx}%
  \BibitemOpen
  \bibfield  {author} {\bibinfo {author} {\bibfnamefont {F.}~\bibnamefont
  {Marchesano}}, \bibinfo {author} {\bibfnamefont {L.}~\bibnamefont
  {Melotti}},\ and\ \bibinfo {author} {\bibfnamefont {L.}~\bibnamefont
  {Paoloni}},\ }\bibfield  {title} {\bibinfo {title} {{On the moduli space
  curvature at infinity}},\ }\href {https://doi.org/10.1007/JHEP02(2024)103}
  {\bibfield  {journal} {\bibinfo  {journal} {JHEP}\ }\textbf {\bibinfo
  {volume} {02}},\ \bibinfo {pages} {103}},\ \Eprint
  {https://arxiv.org/abs/2311.07979} {arXiv:2311.07979 [hep-th]} \BibitemShut
  {NoStop}%
\bibitem [{\citenamefont {Marchesano}\ \emph {et~al.}(2025)\citenamefont
  {Marchesano}, \citenamefont {Melotti},\ and\ \citenamefont
  {Wiesner}}]{Marchesano:2024tod}%
  \BibitemOpen
  \bibfield  {author} {\bibinfo {author} {\bibfnamefont {F.}~\bibnamefont
  {Marchesano}}, \bibinfo {author} {\bibfnamefont {L.}~\bibnamefont
  {Melotti}},\ and\ \bibinfo {author} {\bibfnamefont {M.}~\bibnamefont
  {Wiesner}},\ }\bibfield  {title} {\bibinfo {title} {{Asymptotic curvature
  divergences and non-gravitational theories}},\ }\href
  {https://doi.org/10.1007/JHEP02(2025)151} {\bibfield  {journal} {\bibinfo
  {journal} {JHEP}\ }\textbf {\bibinfo {volume} {02}},\ \bibinfo {pages}
  {151}},\ \Eprint {https://arxiv.org/abs/2409.02991} {arXiv:2409.02991
  [hep-th]} \BibitemShut {NoStop}%
\bibitem [{\citenamefont {Castellano}\ \emph {et~al.}(2024)\citenamefont
  {Castellano}, \citenamefont {Marchesano}, \citenamefont {Melotti},\ and\
  \citenamefont {Paoloni}}]{Castellano:2024gwi}%
  \BibitemOpen
  \bibfield  {author} {\bibinfo {author} {\bibfnamefont {A.}~\bibnamefont
  {Castellano}}, \bibinfo {author} {\bibfnamefont {F.}~\bibnamefont
  {Marchesano}}, \bibinfo {author} {\bibfnamefont {L.}~\bibnamefont
  {Melotti}},\ and\ \bibinfo {author} {\bibfnamefont {L.}~\bibnamefont
  {Paoloni}},\ }\bibfield  {title} {\bibinfo {title} {{The Moduli Space
  Curvature and the Weak Gravity Conjecture}},\ }\href@noop {} {\  (\bibinfo
  {year} {2024})},\ \Eprint {https://arxiv.org/abs/2410.10966}
  {arXiv:2410.10966 [hep-th]} \BibitemShut {NoStop}%
\bibitem [{\citenamefont {Castellano}\ \emph {et~al.}(2026)\citenamefont
  {Castellano}, \citenamefont {Marchesano},\ and\ \citenamefont
  {Paoloni}}]{Castellano:2026bnx}%
  \BibitemOpen
  \bibfield  {author} {\bibinfo {author} {\bibfnamefont {A.}~\bibnamefont
  {Castellano}}, \bibinfo {author} {\bibfnamefont {F.}~\bibnamefont
  {Marchesano}},\ and\ \bibinfo {author} {\bibfnamefont {L.}~\bibnamefont
  {Paoloni}},\ }\bibfield  {title} {\bibinfo {title} {{Curvature divergences
  and gravity decoupling in Calabi--Yau rigid limits}},\ }\href@noop {} {\
  (\bibinfo {year} {2026})},\ \Eprint {https://arxiv.org/abs/2602.04957}
  {arXiv:2602.04957 [hep-th]} \BibitemShut {NoStop}%
\bibitem [{\citenamefont {Enr\'\i{}quez~Rojo}\ and\ \citenamefont
  {Plauschinn}(2020)}]{Enriquez-Rojo:2020pqm}%
  \BibitemOpen
  \bibfield  {author} {\bibinfo {author} {\bibfnamefont {M.}~\bibnamefont
  {Enr\'\i{}quez~Rojo}}\ and\ \bibinfo {author} {\bibfnamefont
  {E.}~\bibnamefont {Plauschinn}},\ }\bibfield  {title} {\bibinfo {title}
  {{Swampland conjectures for type IIB orientifolds with closed-string
  U(1)s}},\ }\href {https://doi.org/10.1007/JHEP07(2020)026} {\bibfield
  {journal} {\bibinfo  {journal} {JHEP}\ }\textbf {\bibinfo {volume} {07}},\
  \bibinfo {pages} {026}},\ \Eprint {https://arxiv.org/abs/2002.04050}
  {arXiv:2002.04050 [hep-th]} \BibitemShut {NoStop}%
\end{thebibliography}%

\end{document}